\documentclass[fleqn,usenatbib]{mnras}

\usepackage[T1]{fontenc}
\usepackage{ae,aecompl}
 
\usepackage{graphicx}	
\usepackage{amsmath}	

\usepackage{amssymb}	

\title[X-ray absorption due to the WHIM]{Searching for the warm-hot intergalactic medium using {\it XMM-Newton} high-resolution X-ray spectra.}

\author[Gatuzz et al.]{
E. Gatuzz$^{1}$\thanks{E-mail: efraingatuzz@gmail.com},
Javier~A.~Garc\'ia$^{2}$, 
E. Churazov$^{3}$,
and T.~R.~Kallman$^{4}$,
\\
$^{1}$Max-Planck-Institut f\"ur extraterrestrische Physik, Gie{\ss}enbachstra{\ss}e 1, 85748 Garching, Germany\\
$^{2}$Cahill Center for Astronomy and Astrophysics, California Institute of Technology, Pasadena, CA 91125, USA\\
$^{3}$Max-Planck-Institut f\"ur Astrophysik, D-85741 Garching bei M\"unchen, Germany\\  
$^{4}$NASA Goddard Space Flight Center, Greenbelt, MD 20771, USA\\ 
}

\date{Accepted XXX. Received YYY; in original form ZZZ} 
\pubyear{2020} 
\begin{document}
 \label{firstpage}
\pagerange{\pageref{firstpage}--\pageref{lastpage}}
\maketitle 
\begin{abstract} 

The problem of missing baryons in the local universe remains an open question. One propose alternative is that at low redshift missing baryons are in the form of the Warm Hot Intergalactic Medium (WHIM).  In order to test this idea,  we present a detailed analysis of X-ray high-resolution spectra of six extragalactic sources, Mrk~421, 1ES~1028+511, 1ES~1553+113, H2356-309, PKS~0558-504 and PG~1116+215, obtained with the {\it XMM-Newton} Reflection Grating Spectrometer to search for signals of WHIM  and/or circumgalactic medium (CGM) X-ray absorbing gas. We fit the X-ray absorption with the {\tt IONeq} model, allowing us to take into account the presence of X-ray spectral features due to the multiphase component of the local ISM. An additional {\tt IONeq} component is included to model the WHIM absorption, instead of the traditional Gaussian absorption line modeling. We found no statistical improvement in the fits when including such component in any of the sources, concluding that we can safely reject a successful detection of WHIM absorbers towards these lines of sights. Our simulation shows that the presence of the multiphase ISM absorption features prevents detection of low-redshift WHIM absorption features in the $>17$~\AA\ spectral region for moderate exposures using high-resolution spectra.
\end{abstract}


 \section{Introduction }\label{sec_int}
Matter detected in form of interstellar medium (ISM), stars, galaxies and intracluster medium adds up to 30-40$\%$ of the estimated baryonic density in the nearby universe \citep{shu12,yao12}. Cosmological simulations indicate that most of the ``missing baryons'' reside in the form of WHIM, filamentary hot structures among galaxies with temperatures T $\sim 10^{5}-10^{7}$ K \citep{dav01}. Absorption lines due to high-ionization states (e.g. {\rm O}~{\sc vii} and {\rm O}~{\sc viii}) detected in the Intergalactic Medium (IGM) are good indicators of the presence of the WHIM \citep{nel18}. 
  
Due to the weak X-ray signal of the WHIM and the limitations of the effective area of the current X-ray observatories, statistically significant detection of absorption lines due to the WHIM embedded within the large-scale filament structures are very sparse. By using an extragalactic bright X-ray source, which acts as a lamp, the WHIM absorber could be detected. Using {\it Chandra} X-ray high-resolution spectra,  \citet{nic05a,nic05b} reported the first 3$\sigma$ detection of absorption systems associated to WHIM gas in the line of sight to the blazar Mkn~421, at $z=0.027$ and $z=0.011$. However, this result was questioned by \citet{kaa06} and \citet{ras07} based on statistical arguments, apparent wavelength inconsistencies for some of the lines and the lack of detection of such systems in {\it XMM-Newton} RGS spectra. Claims about the identification of WHIM absorption lines in the Sculptor Wall have been done by \citet{buo09, fan10, zap10} by analyzing X-ray high-resolution spectra of the blazar H2356-309, which were called into question by \citet{wil13a} due to the existence of a galaxy 240 kpc away from the absorber. More important, it has been shown by \citet{nic16b} that such absorption lines could be contaminated by Galactic {\rm O}~{\sc ii} K$\beta$ absorption. That is, the inclusion of the multiphase ISM component in the modeling of X-ray absorption spectral features is crucial to avoid confusion between Galactic, external and intrinsic to the source features when interpreting the results \citep[see for example][]{gat20}. 
 
In the last decade, an enormous effort has been done not only in the computation of photoabsorption cross-sections with the state-of-the art techniques \citep{gar05,gar09a,wit09,wit11a,wit11b,gor13,pal16} but also in the benchmarking of the atomic data using astronomical observations and experimental measurements, including species such as oxygen \citep{gat13a,gat13b,gor13,gat14,leu20}, neon \citep{gat15}, magnesium \citep{has14}, carbon \citep{gat18c}, silicon \citep{gat20b} and nitrogen \citep{gat21a}. Such benchmarking of the atomic data constitute a crucial step. For example, \citet{gat13a} show a disagreement between the {\it Chandra} K$\alpha$ absorption line position and the experimental measurements by \citet{sto97}. Uncertainties about the experimental energy calibration scale suggested that the astrophysical measurement was correct, which was confirmed by the laboratory measurements done by \citet{leu20}.
 
Here, we present the analysis of {\it XMM-Newton} X-ray high-resolution spectra of six extragalactic sources for which WHIM absorbers have been identified in the past (i.e. Mrk~421,1ES~1028+511, 1ES~1553+113, H2356-309, PG~1116+215 and PKS~0558-504), searching for WHIM spectral absorption features. The outline of the present paper is as follows. In Section~\ref{sec_dat}, we described the sources analyzed, including previous claims of WHIM spectral features detection and the data reduction aspects of the observations. Section~\ref{sec_model} describes the spectral fitting procedure, followed by a detailed discussion of the best-fit results in Section~\ref{sec_disc}. Finally, we draw in Section~\ref{sec_con} the conclusions of this work.   
  
\begin{table*}
\small
\caption{\label{tab_obs}Observations analyzed in the present work.}
\centering
\begin{tabular}{lccccccc}
\hline
Source&Galactic& Redshift & $N({\rm H})$-21~cm$^{a}$ &$\#$ Obs.&  Filtered exposure    \\
Name&coordinates& & &&time (Ms)\\
\hline
Mrk~421      & (166.05,38.19)  & 0.031& 2.01& 38& 2.04\\
1ES~1028+511 & (157.70,50.85)  & 0.361& 1.26& 4&  0.61\\  
1ES~1553+113 & (238.99,11.17)  & 0.490& 4.35& 22& 3.95\\
H2356-309    & (359.87,-30.59) & 0.165& 1.48& 8&  1.11\\
PKS~0558-504 & (89.84,-50.45)  & 0.137& 4.18& 13& 1.73\\   
PG~1116+215  & (169.75,21.23)  & 0.177& 1.43& 6&  0.78\\   
\hline 
\multicolumn{6}{l}{ $^{a}$Column densities obtained from \citet{wil13} and are in units of $10^{20}$ cm$^{-2}$.}\\
\end{tabular}
\end{table*}  

\section{Observations and data reduction}\label{sec_dat}
We have analyzed the following six sources for which WHIM absorbers have been identified in the past using X-ray observations:
\begin{itemize}
\item[--] {\bf Mrk~421:} is a blazar located at z = 0.031 \citep{cro07}. By analyzing {\it Chandra} observations, \citet{nic05b} identified an absorber located at $z=0.027$, with statistical significance of 4.9$\sigma$. The absorber is traced mainly by {\rm O}~{\sc vii} and {\rm N}~{\sc vii}. However, \citet{nic16b} claimed that such absorption lines are contaminated by Galactic {\rm O}~{\sc ii} K$\beta$ absorption.
\item[--] {\bf 1ES~1028+511:} is a high energy peaked BL Lac located at $z = 0.361$ \citep{don01}. The optical R band photometry shows significant lightcurve variations \citep{kur04}. Using {\it Chandra} observations, \citet{nic05} identified two absorption lines at $28.740\pm 0.038$\AA\ and $48.816\pm 0.040$\AA\, which were associated to two WHIM systems traced by {\rm O}~{\sc vii} and {\rm C}~{\sc v}, respectively, and located at $z=0.330\pm 0.002$ and $z=0.212\pm 0.001$. \citet{ste06}, using both {\it XMM-Newton} and {\it Chandra} X-ray spectra, identified three absorption lines, identified as {\rm O}~{\sc vii}, {\rm C}~{\sc iv} and {\rm C}~{\sc v}, and associated to WHIM systems located in the $z=0.12-0.21$ range. 
\item[--] {\bf H2356-309:} is a Blazar located at $z=0.165$ \citep{don01}. The source is located behind the the Sculptor Wall \citep[$z=0.028-0.032$,][]{dac94} and Pisces-Cetus \citep[$z=0.0545-0.0625$,][]{por05} superclusters, making it a candidate to study WHIM absorber in its line-of-sight. Using X-ray high-resolution spectra \citet{buo09,fan10,nev15} reported the detection of an {\rm O}~{\sc vii} absorption line at $z=0.0311-0.0327$ associated to the Sculptor Wall. The detection was confirmed by \citet{zap10}, who also identified a {\rm C}~{\sc v} absorption line at $z=0.127-0.129$. However, \citet{will13} suggested that the absorbers trace gas associated to nearby galaxies rather than filamentary structures.
\item[--] {\bf PKS~0558-504:} is a Seyfert galaxy located at $z=0.1379$ \citep{era04}. The properties of the AGN itself has been extensively studied \citep[e.g.][]{gli10,gli13,han14,gho16}.  \citet{nic10} reported a $<3\sigma $ detection of an {\rm O}~{\sc viii} absorption line at $z=0.117\pm 0.001$ associated to the WHIM using X-ray spectra from {\it XMM-Newton} in combination with far-UV {\it FUSE} data. They pointed out that, based on X-ray data only, the parameters of the model are poorly constrained. 
\item[--] {\bf 1ES~1553+113:} is a Blazar located at $z=0.49\pm 0.04$ \citep{hes15}. \citet{nic18} reported the detection of two {\rm O}~{\sc vii} absorbers associated to the WHIM at $z=0.4339\pm 0.0008$ and $z=0.3551^{+0.0003}_{-0.0015}$. On the other hand, \citet{joh19} concluded in their analysis of UV and X-ray spectra that such absorber arises from the intergalactic medium surrounding the galaxy instead of the WHIM.
\item[--] {\bf PG~1116+215:} is a quasar located at $z=0.177$ \citep{til12} and well studied in the UV band \citep{sem04,leh07,til12}. Using {\it Chandra} and {\it XMM-Newton} X-ray spectra, \citet{bon16} identified an {\rm O}~{\sc viii} absorption line associated to the WHIM at $z=0.0911\pm 0.0009$, which may correspond to a galaxy filament in the line of sight to the source \citep{tem14}. A {\rm O}~{\sc vii} $K\alpha$ absorption line, which would correspond to the same absorber, is marginally detected.
\end{itemize}

Table~\ref{tab_obs} list the specifications for all observations analyzed, including column densities obtained from \citet{wil13}. RGS spectra reduction, including background substraction, was done following the Scientific Analysis System (SAS, version 18.0.0) threads\footnote{\url http://xmm.esac.esa.int/sas/current/documentation/threads/}. The specific {\it XMM-Newton} observation IDs considered in our sample are listed in Appendix~\ref{sec_obsids}. For each source, we have selected spectra with $>1000$ counts in the spectral fitting range (8--35 \AA). We used the most up-to-date effective area corrections by switching on the parameters {\tt withrectification} and {\tt witheffectiveareacorrection} of the SAS {\tt rgsproc} tool that generates the RGS response matrices. To make sure that the extraction mask is not far off the source position we manually entered the source celestial coordinates in the processing script. For each observation, we created light-curves and then removed periods of high background to create cleaned event files. First-order source and background spectra were produced from these cleaned events, together with response matrices. The spectra of each observation were background subtracted. For each source, all observations were combined using the {\tt rgscombine} task in order to maximize the Signal to Noise per Resolution Element (SNRE). The spectral fitting was carried out with the {\sc xspec} analysis data package \citep[version 12.10.1\footnote{\url{https://heasarc.gsfc.nasa.gov/xanadu/xspec/}}]{arn96}. All spectra were rebinned to 1 count per channel in combination with the \citet{chu96} weighting method which allows the analysis of low-count spectra in combination with $\chi^{2}$ statistics \citep[see for example,][]{tof13,joa16,mil16d,gat18,med18}. Finally, the abundances are given relative to \citet{gre98}.

 \section{X-ray spectra modeling}\label{sec_model}
  
 \subsection{The ISM X-ray absorption component}\label{sec_model_ism}
  
\begin{figure}  
\includegraphics[scale=0.42]{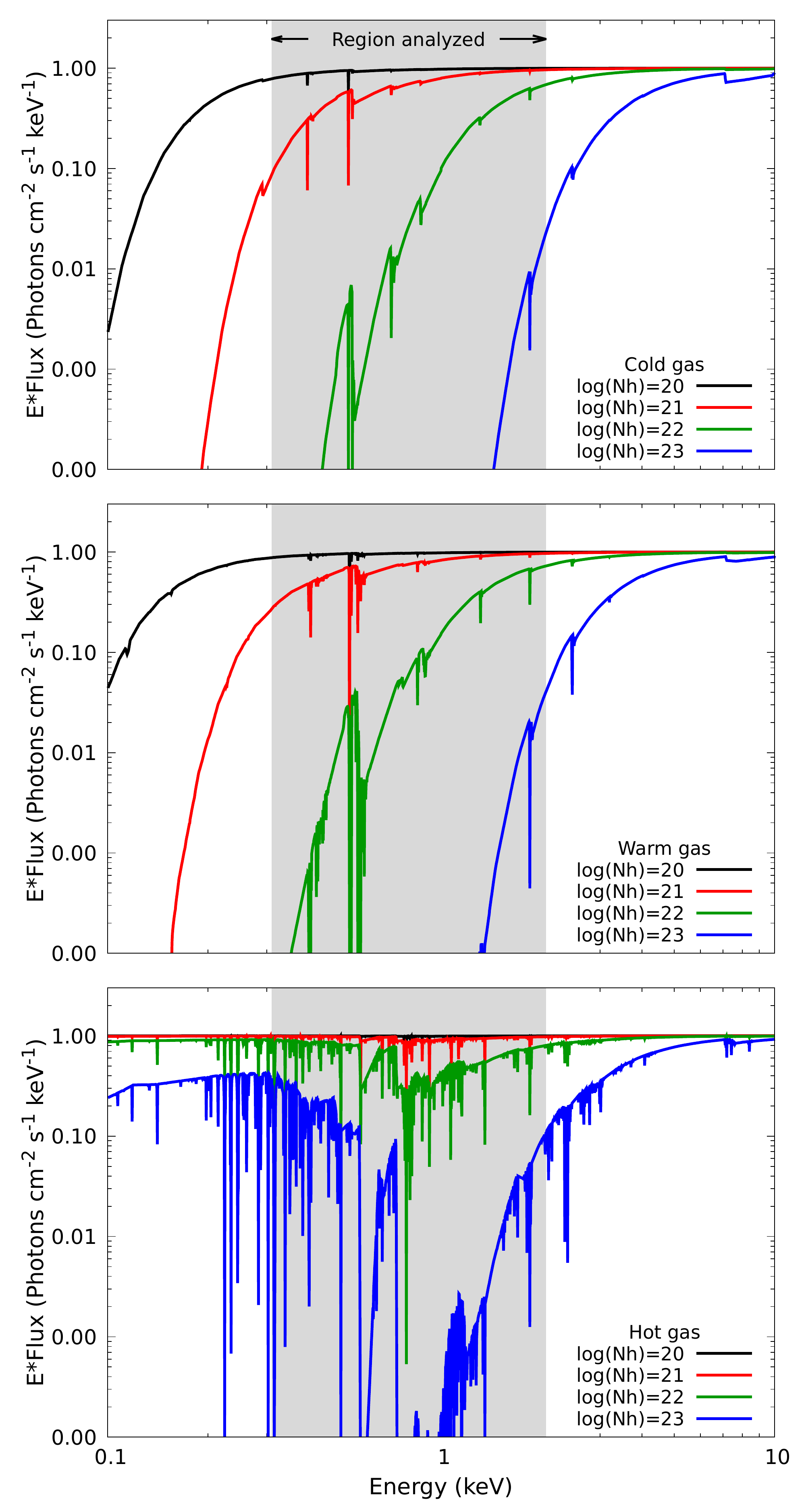}     
\caption{Effects of the $N({\rm H})$ column densities in the {\tt IONeq} model for the cold, warm and hot components. The shaded regions indicates the bandpass analyzed (see Section~\ref{sec_model}).} \label{fig_ioneq_model} 
\end{figure}

In order to model the X-ray absorption due to the local ISM we use the {\tt IONeq} model \citep{gat18}. {\tt IONeq} assumes collisional ionization equilibrium (CIE) and includes the hydrogen column density ($N({\rm H})$), gas temperature ($T_{e}$), oxygen, neon and iron abundances ($A_{x}$), redshift and turbulence broadening ($v_{turb}$) as parameters of the model\footnote{For a detailed description on the atomic data included in the model see \citet{gat18}}. We fit the contribution of the ISM for each source with the following model (using {\sc xspec} nomenclature):
\begin{center}
{\tt IONeq}$_{cold}$*{\tt IONeq}$_{warm}$*{\tt IONeq}$_{hot}$*{\tt powerlaw}
\end{center}
 
where cold, warm and hot refers to the multiphase ISM components. Following the analysis of the ISM from \citet{gat18}, we fixed the gas temperatures to be $T_{\rm{cold}}= 1\times 10^{4}$ K (traced by species such as {\rm N}~{\sc i}, {\rm O}~{\sc i} and {\rm Ne}~{\sc i}), $T_{\rm{warm}}= 5\times 10^{4}$ K (traced by species such as {\rm N}~{\sc ii}, {\rm N}~{\sc iii}, {\rm O}~{\sc ii}, {\rm O}~{\sc iii}, {\rm Ne}~{\sc ii} and {\rm Ne}~{\sc iii}) and $T_{\rm{hot}}= 2\times 10^{6}$ K (traced by species such as {\rm N}~{\sc vi}, {\rm O}~{\sc vii} and {\rm Ne}~{\sc ix}). The abundances were set to solar. For the continuum, and considering the small energy windows analyzed, we use a simple {\tt powerlaw} component. Figure~\ref{fig_ioneq_model} shows the {\tt IONeq} model for different values of $N({\rm H})$ (with a power-law continuum with $\Gamma=2$). Top, middle and bottom panel corresponds to the cold, warm and hot components. The shaded region marks the bandpass analyzed  which is mostly sensitive to columns of $N({\rm H})=10^{20}-10^{23}$~cm$^{-2}$. In this analysis, we assumed $v_{turb}=50$ km s$^{-1}$ for the cold-warm ISM component and $v_{turb}=110$ km s$^{-1}$ for the hot ISM component \citep[i.e. the values obtained by ][]{gat18}.

\begin{figure*}  
\includegraphics[scale=0.6]{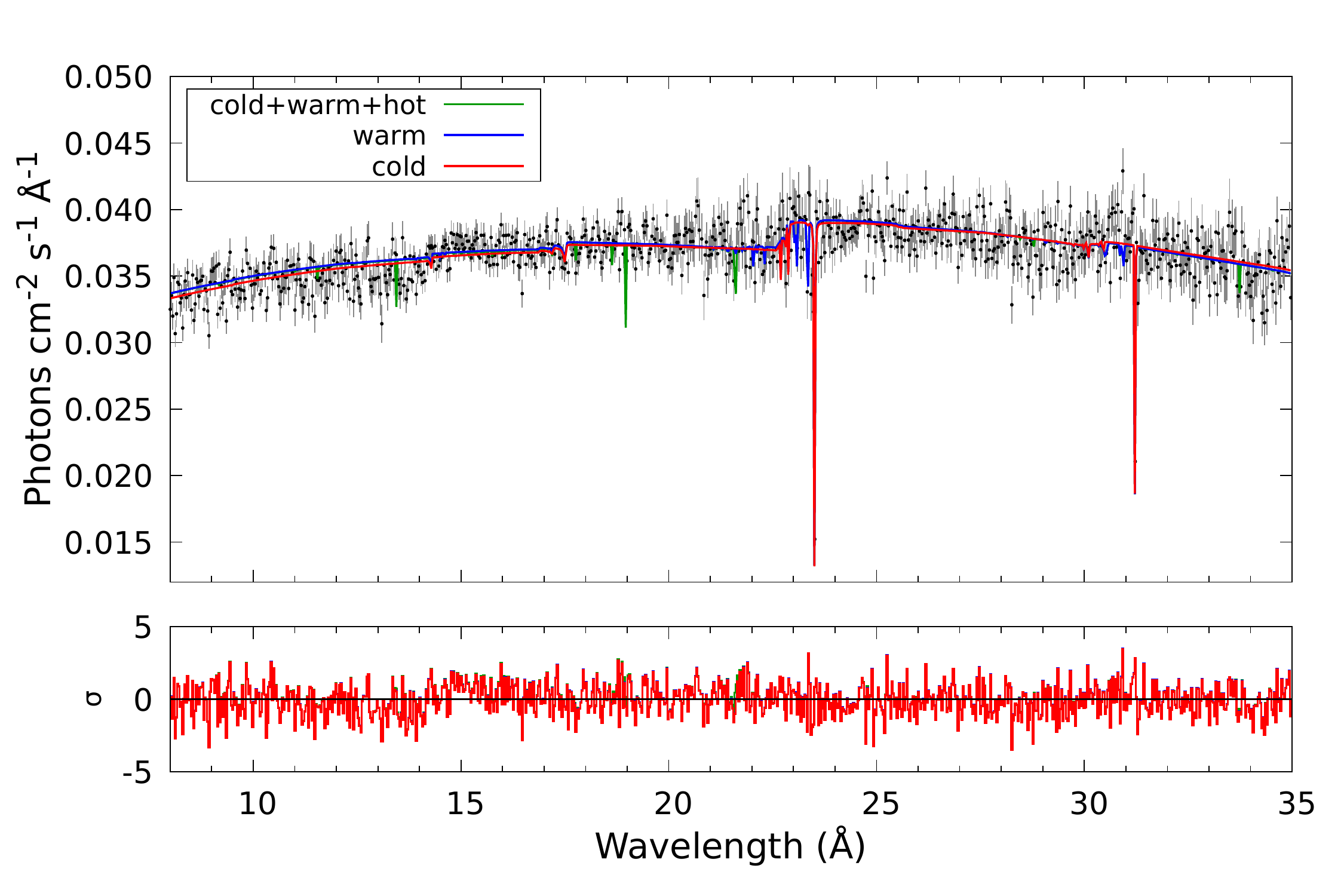}   
\caption{Best-fit results in the 8--35 \AA\ wavelength region for {\it XMM-Newton} Mrk~421. Black data points are the observations, in flux units. Residuals correspond to the $(data-model)/error$ values. Solid line colors indicate the best-fit listed in Table~\ref{tab_best_fit_ism}. The cold and cold-warm models are underneath the cold-warm-hot model. } \label{fig_fits_mrk421A} 
\end{figure*}

\begin{figure}  
\includegraphics[width=0.43\textwidth]{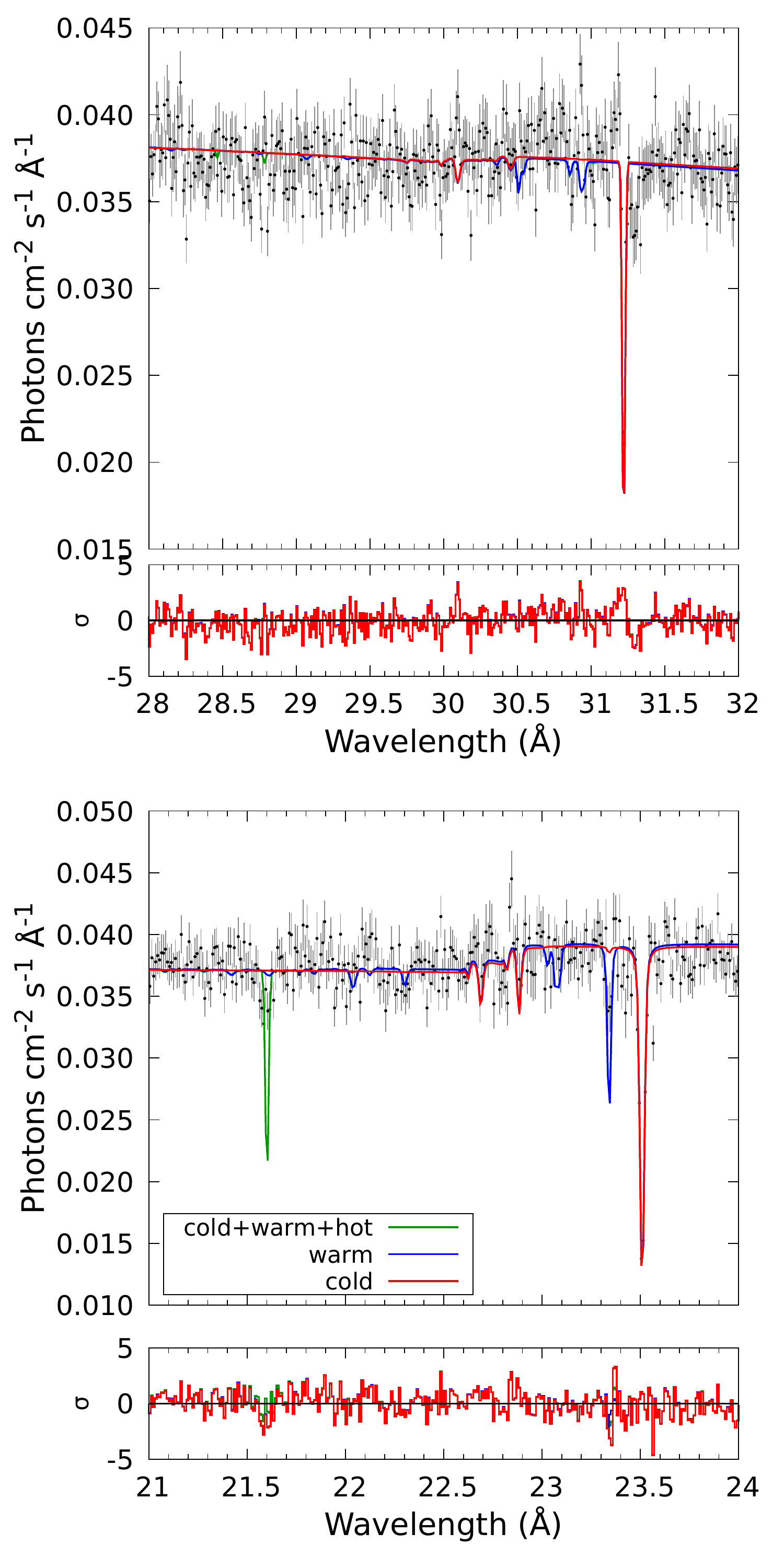}   
\caption{Best-fit results in the N K-edge (top panel) and O K-edge (bottom panel) wavelength region for {\it XMM-Newton} Mrk~421. Solid line colors indicate the best-fit listed in Table~\ref{tab_best_fit_ism}. } \label{fig_fits_mrk421B} 
\end{figure}

Table~\ref{tab_best_fit_ism} shows the best-fit results obtained for all sources. Figure~\ref{fig_fits_mrk421A} shows the best-fit spectra, while Figure~\ref{fig_fits_mrk421B} shows the N and O K-edge wavelength regions for the Mrk~421 source. We choose to show this source in more detail because of its high statistics. We have found that when modeling only the ISM contribution (i.e. cold, warm and hot components) a good fit is achieved for all sources as indicated by the statistical $\chi^{2}$/d.o.f. values\footnote{It should be noted that $\chi^{2}$/d.o.f. values can be biased low in cases when the number of counts per bin is small.} 

\begin{figure*}  
  \includegraphics[scale=0.40]{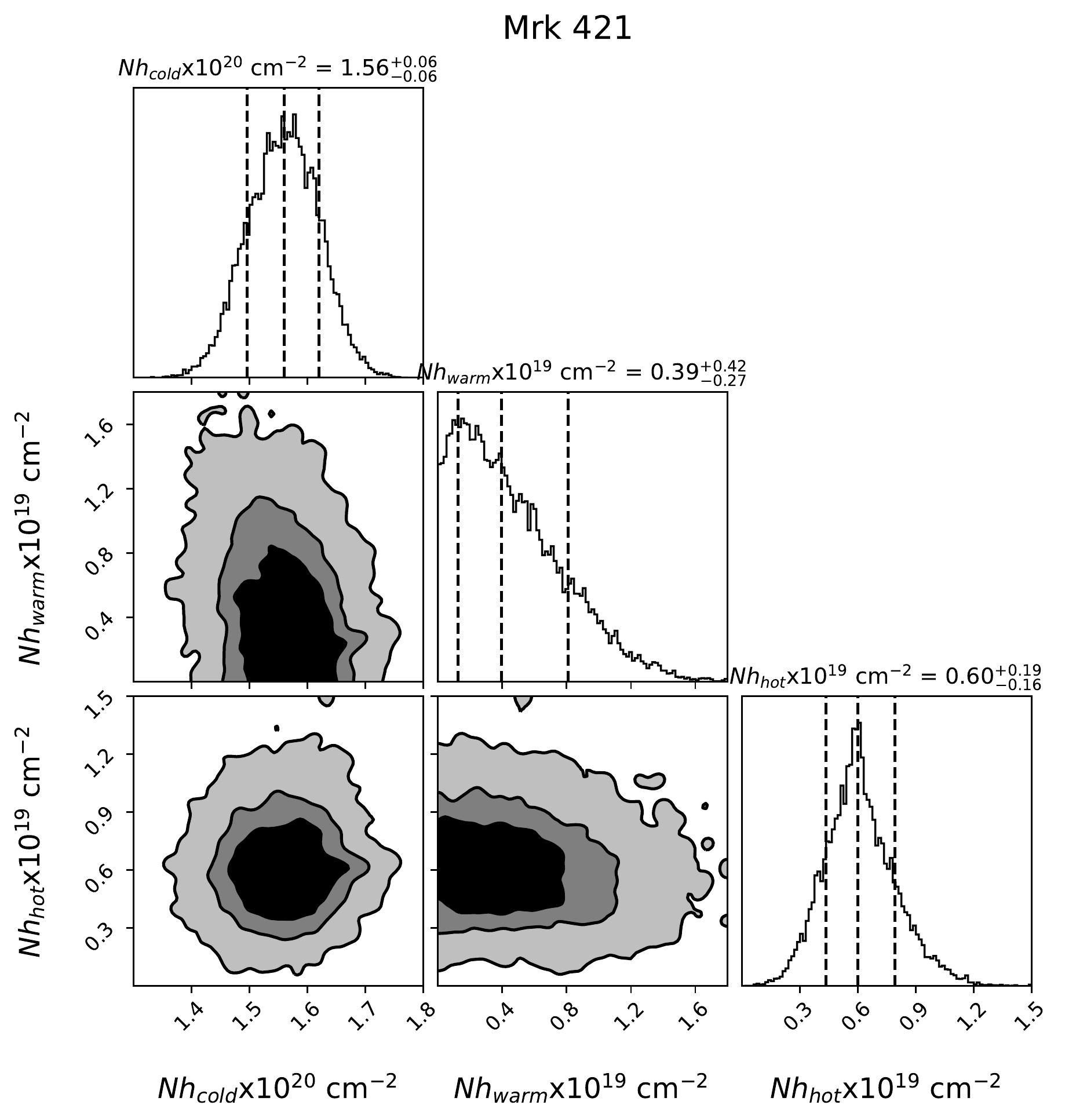}
\includegraphics[scale=0.40]{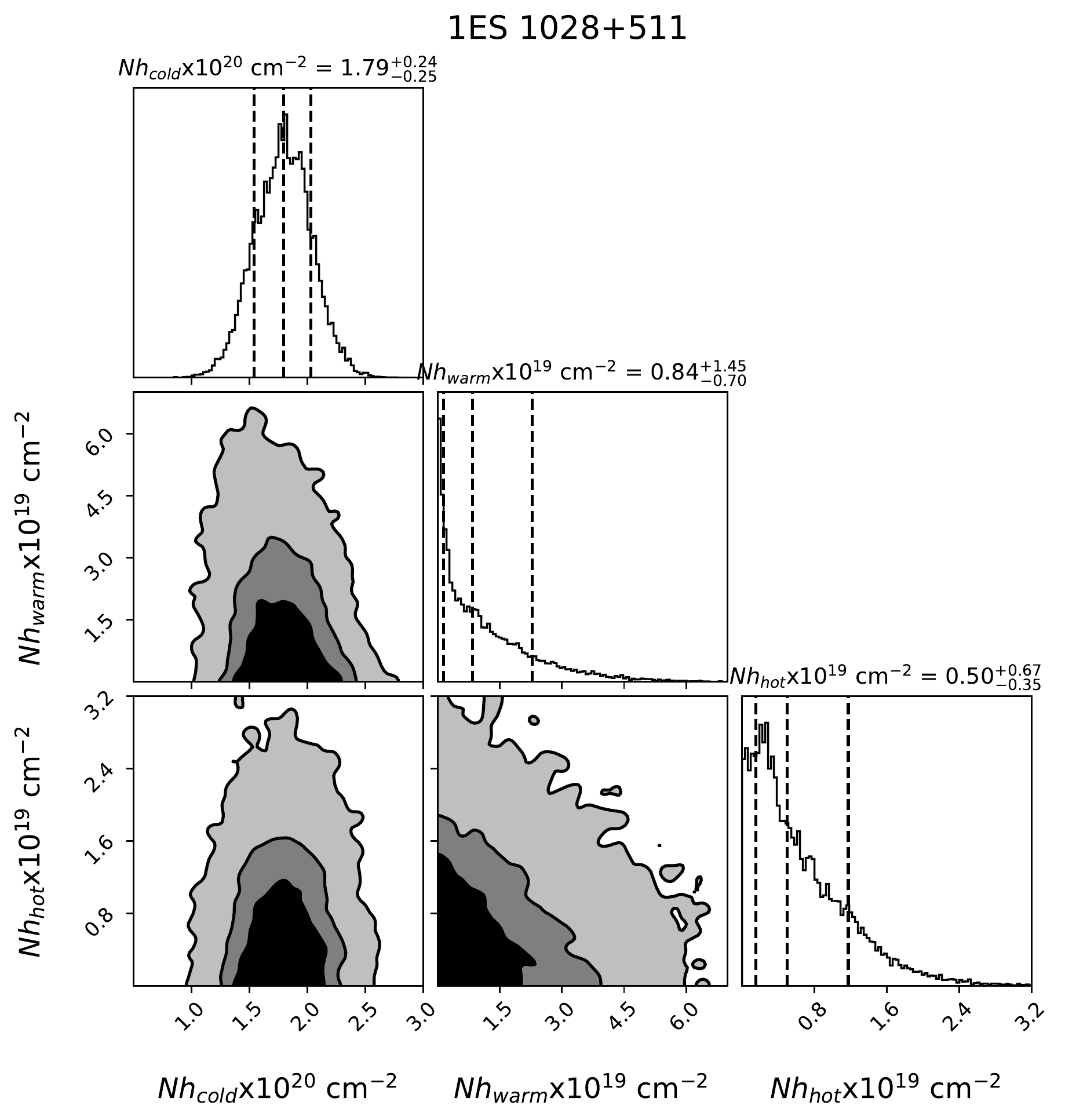}\\ 
  \includegraphics[scale=0.40]{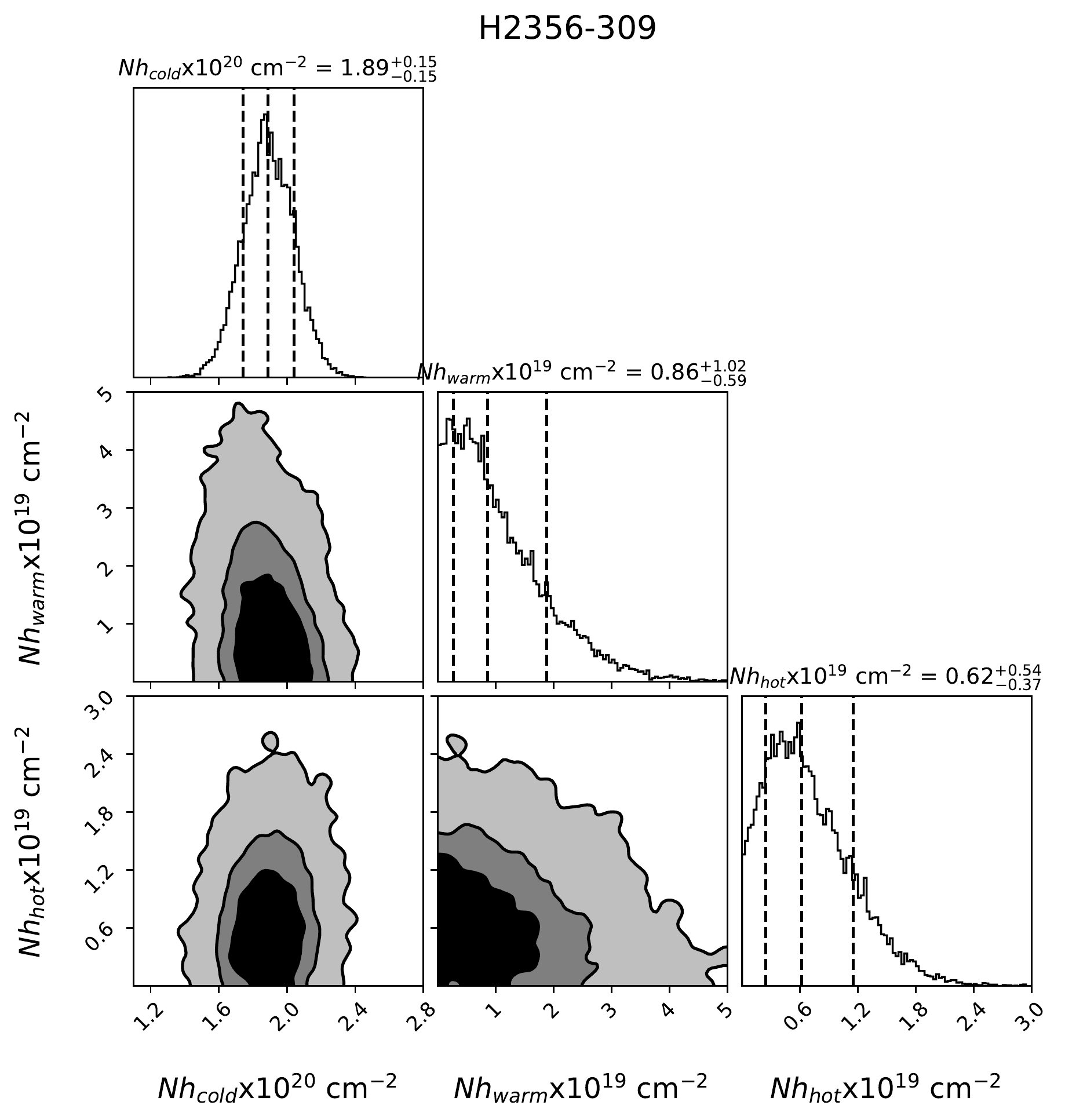}
\includegraphics[scale=0.40]{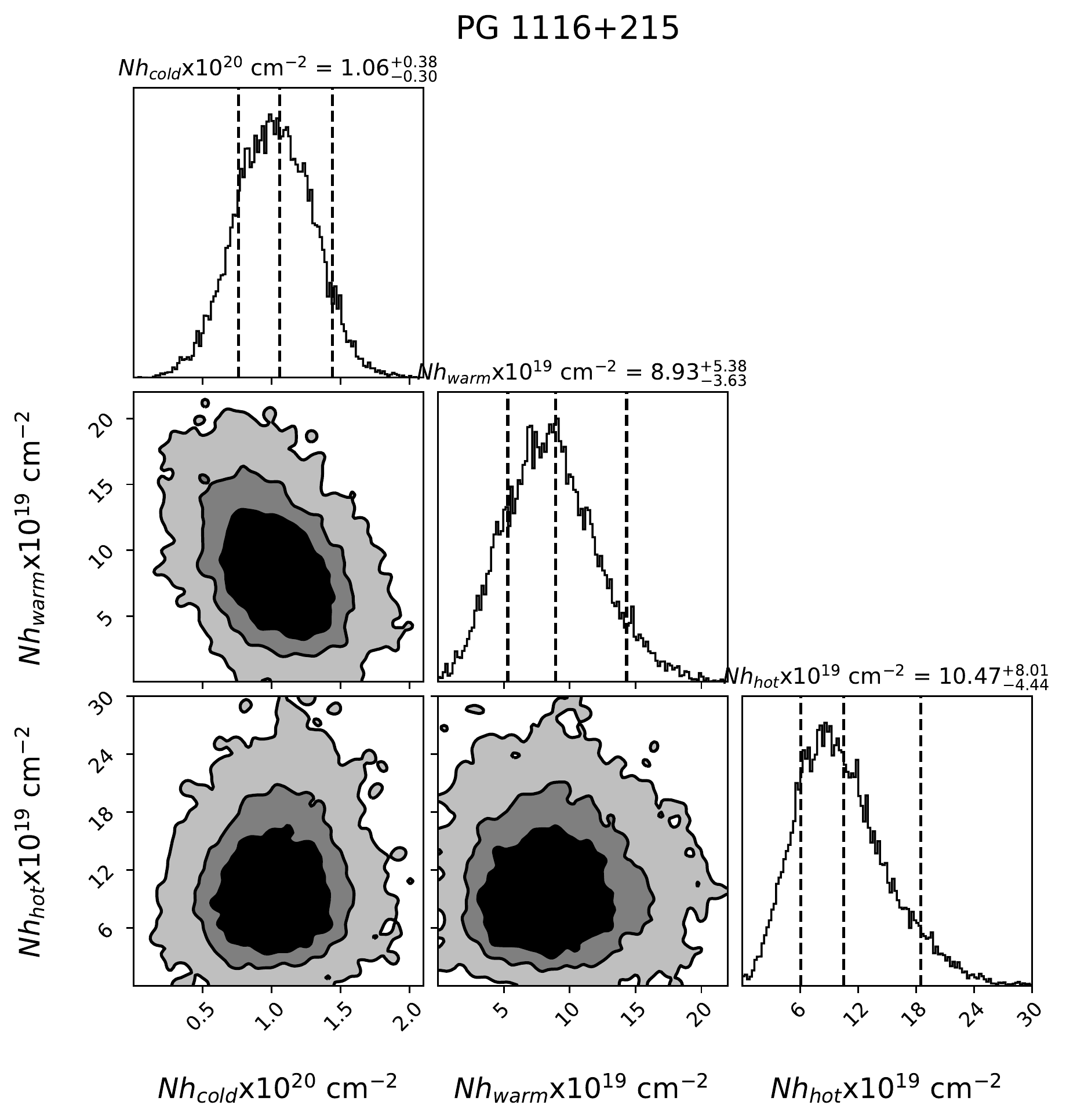}\\ 
\includegraphics[scale=0.40]{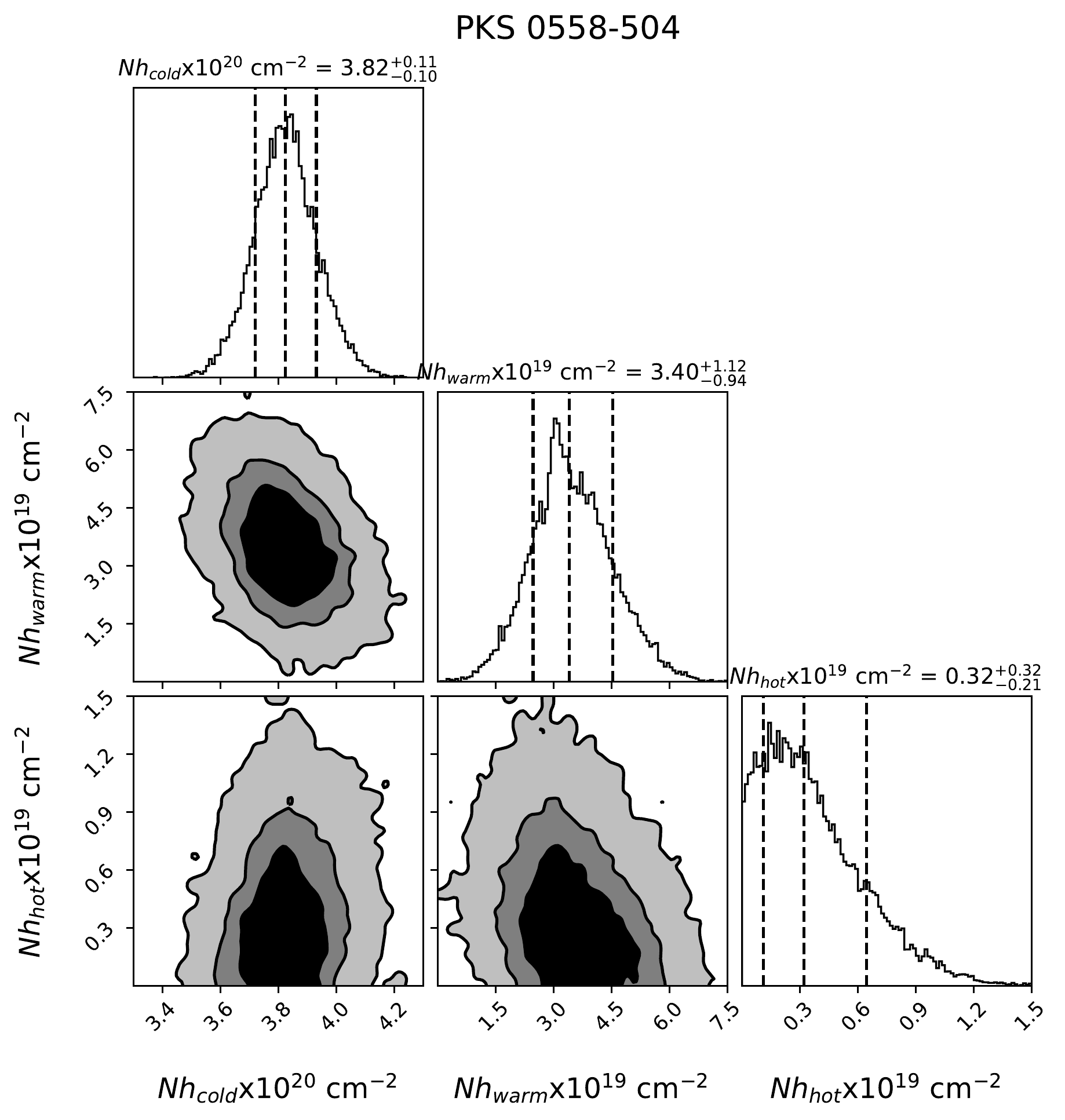}
\includegraphics[scale=0.40]{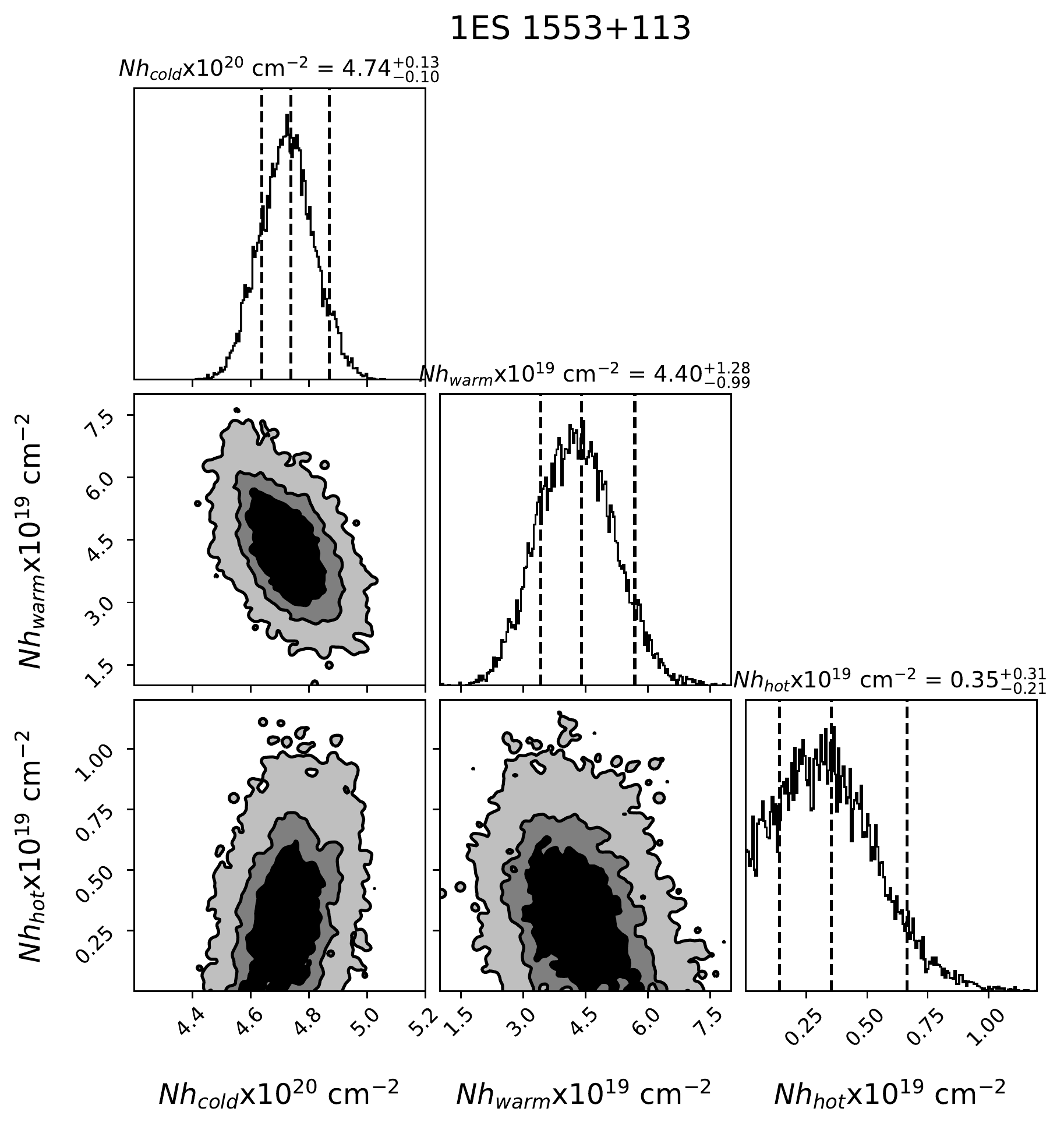}\\
\caption{Analysis of the parameters space for the ISM components. Contour levels correspond to $68\%$, $90\%$, $95\%$ and $99\%$ levels. (see Section~\ref{sec_best_fit_all})} \label{fig_mcmc_par} 
\end{figure*}

For each source we explored the full parameter space for the ISM modeling by using the Markov-chain Monte Carlo (MCMC) method with a Goodman-Weare algorithm. The length of the MCMC chain was 10$^{5}$ with the first 10000 elements corresponding to the burn-in period (i.e. were discarded). Figure~\ref{fig_mcmc_par} show the parameter estimation.  The histograms on the diagonal show the marginalized posterior densities for each parameter while vertical dashed lines represent the 16$^{th}$ and 84$^{th}$ percentiles (see Table~\ref{tab_best_fit_ism_mcmc}).

 Discrepancies between 21-cm measurements and the best-fitting values obtained for the cold ISM component (see Tables~\ref{tab_obs} and~\ref{tab_best_fit_ism}) may be due to the continuum modeling \citep[see ][]{gat18}. Even then, the larger $N({\rm H})$-cold obtained correspond to sources with the larger 21-cm values. The warm+hot phase contribution to the total ISM column density is $<$10$\%$, in agreement with the ratios found in the analysis of extragalactic sources by \citet{gat18}, except for PG~1116+215 where the uncertainties for the warm-hot component are large.

\begin{figure*}  
\includegraphics[scale=0.35]{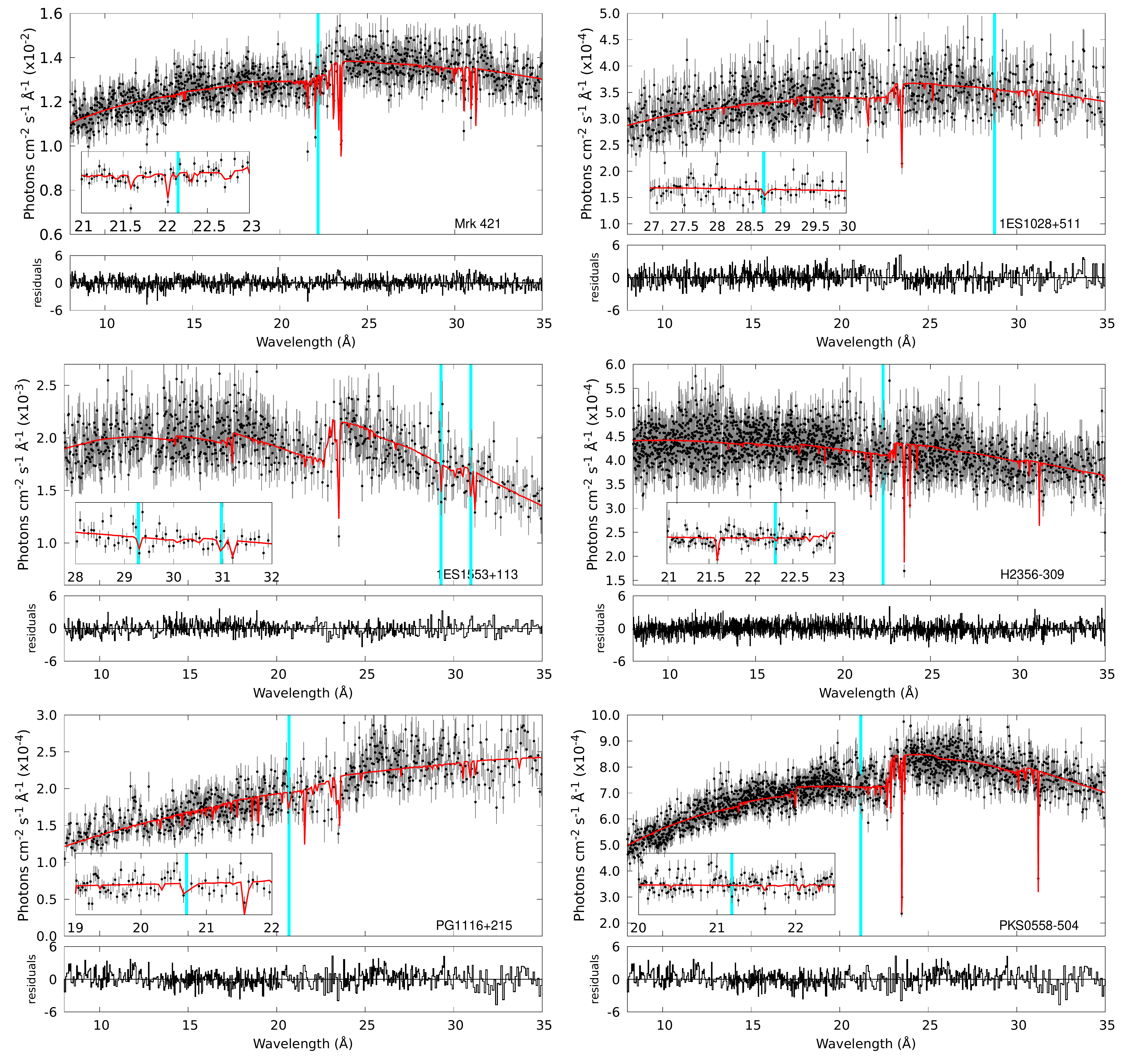}     
\caption{Best-fit results in the 8--35 \AA\ wavelength region for all sources analyzed in flux units. In each panel, the black data points are the observations, while the solid red lines correspond to the best-fit using the {\tt IONeq} models (see Table~\ref{tab_best_fit_ism}). Residuals correspond to the $(data-model)/error$ values. The vertical lines indicate the position were WHIM absorption features should be identified, according to results from previous works. } \label{fig_fits_flux} 
\end{figure*}

Figure~\ref{fig_fits_flux} shows the best-fit spectra in flux units for this model. For illustrative purposes, spectra from multiple observations have been combined and rebinned to have at least 10 counts per channel. Vertical dashed lines indicate the position were WHIM absorption features have been identified in previous works (see descriptions in Section~\ref{sec_dat}). Our best-fit results show that such WHIM identified spectral fingerprints may be misidentified given their presence close to absorption features due to the ISM. Finally, we performed a simultaneous fit of all observations for each source in order to take into account variations in the continuum. We found that the best-fit parameters of the ISM multicomponent are in good agreement with values shown in Table~\ref{tab_best_fit_ism}, thus verifying the validity of the combined data fit approach. The worst fit is obtained for 1ES~1553+113 for which we identified large residuals in the continuum in regions beyond the absorption N K-edge as well as multiple bad pixels that are not removed by the standard reduction scripts. Such additional cool pixels in the dispersing detector increase the uncertainties in the RGS effective area and were also identified by \citep{nic18}. We tested the best-fitting results by modeling the data including only wavelength regions around the main absorption edges (i.e. $13-15$~\AA for the Ne K-edge, $16-18$~\AA for the metallic iron, $18-24$~\AA for the O K-edge and the $28-31.5$~\AA for the N K-edge). The column densities are in good agreement with results listed in Table~\ref{tab_best_fit_ism} while the best-fit statistic is significantly better (cstat/d.o.f.$=1308/1135=1.15$).

\begin{table*}
\small
\caption{Best-fit parameters for the ISM component.}\label{tab_best_fit_ism}
\begin{tabular}{llccccccccccccccccccc}
\hline
Model&Parameter&1ES~1028+511& 1ES~1553+113  & H2356-309 & Mrk~421 & PG~1116+215 & PKS~0558-504   \\ 
\hline
 {\tt IONeq}$^{a}$ & $N({\rm H})$-cold &$1.81^{+0.12}_{-0.25}$  &$ 4.78  \pm 0.08  $  &$ 1.91 \pm 0.15 $  &$ 1.61 \pm 0.04  $    &$ 0.94 \pm 0.27 $  &$  3.88_{-0.10}^{+0.17}  $     \\
&$N({\rm H})$-warm &$ <0.83 $  &$3.78_{-0.77}^{+0.85}  $  &$ <1.15 $  &$ 0.17_{-0.08}^{+0.04}   $    &$ 8.74_{-3.33}^{+3.85}  $  &$ 3.12_{-1.01}^{+0.79} $     \\
&$N({\rm H})$-hot &$ <0.97  $  &$  0.37_{-0.20}^{+0.22}  $  &$ 0.57_{-0.46}^{+0.54}$  &$0.59_{-0.01}^{+0.08}   $    &$  8.61_{-3.91}^{+5.02}  $  &$  0.33_{-0.25}^{+0.31} $     \\
{\tt Powerlaw}$^{b}$&$\Gamma$ &$2.29 \pm 0.02 $  &$ 2.54 \pm 0.01 $  &$ 2.07\pm 0.01 $  &$ 2.21 \pm 0.01  $    &$ 2.61 \pm 0.03 $  &$  2.67\pm 0.01$     \\
&$norm$ &$4.05 \pm 0.03 $  &$ 6.35 \pm 0.02  $  &$  5.65\pm 0.02  $  &$ 459 \pm 1  $    &$ 1.95 \pm 0.02 $  &$8.41 \pm 0.02$     \\
statistics &cstat/d.o.f. &$ 2737/2660  $  &$  3769/2694  $  &$  2813/2691  $  &$  3089/2686  $    &$  2807/2669  $  &$  2948/2694 $     \\
&red-$\chi^{2}$&$ 1.03$  &$  1.39   $  &$ 1.04   $  &$ 1.15    $    &$  1.05   $  &$ 1.09   $     \\ 
\hline
\multicolumn{7}{l}{$^{a}$ $N({\rm H})$-cold in units of $10^{20}$ cm$^{-2}$. $N({\rm H})$-warm and $N({\rm H})$-hot in units of $10^{19}$ cm$^{-2}$. }\\
\multicolumn{7}{l}{ $^{b}$ Power-law normalization is in units of $10^{-3}$ photons keV$^{-1}$ cm$^{-2}$ s$^{-1}$ at 1~keV } 
\end{tabular}
\end{table*}

\begin{table*}
\small
\caption{16$^{th}$ and 84$^{th}$ percentiles for the ISM component obtained from the MCMC analysis. }\label{tab_best_fit_ism_mcmc}
\begin{tabular}{llccccccccccccccccccc}
\hline
Model&Parameter&1ES~1028+511& 1ES~1553+113  & H2356-309 & Mrk~421 & PG~1116+215 & PKS~0558-504   \\ 
\hline
 {\tt IONeq}$^{a}$ & $N({\rm H})$-cold &$1.79\pm 0.25  $  &$4.74_{-0.10}^{+0.13}   $  &$1.89\pm 0.15  $  &$1.56\pm 0.06   $    &$1.06_{-0.30}^{+0.38}   $  &$3.82\pm 0.10$     \\
&$N({\rm H})$-warm &$0.84_{-0.70}^{+1.45} $  &$4.40_{-0.99}^{+1.28} $  &$0.86_{-0.59}^{+1.02}  $  &$ 0.39_{-0.27}^{+0.42}  $    &$ 8.93_{-3.63}^{+5.38}  $  &$3.40_{-0.94}^{+1.12} $     \\
&$N({\rm H})$-hot &$0.50_{-0.35}^{+0.67} $  &$0.35_{-0.21}^{+0.31}$  &$ 0.62_{-0.37}^{+0.54} $  &$0.60_{-0.16}^{+0.19}   $    &$ 10.47_{-4.44}^{+8.01}  $  &$0.32_{-0.21}^{+0.32} $     \\
{\tt Powerlaw}$^{b}$&$\Gamma$ &$2.29\pm 0.02 $  &$ 2.54\pm 0.01 $  &$2.07\pm 0.01  $  &$ 2.21\pm 0.01  $    &$2.62\pm 0.03   $  &$2.66\pm 0.01 $     \\
&$norm$ &$4.06\pm 0.29 $  &$ 6.35\pm 0.02 $  &$ 5.66\pm 0.02 $  &$459.04\pm 0.76   $    &$ 1.95\pm 0.03  $  &$ 8.41\pm 0.26  $     \\
\hline
\multicolumn{7}{l}{$^{a}$ $N({\rm H})$-cold in units of $10^{20}$ cm$^{-2}$. $N({\rm H})$-warm and $N({\rm H})$-hot in units of $10^{19}$ cm$^{-2}$. }\\
\multicolumn{7}{l}{ $^{b}$ Power-law normalization is in units of $10^{-3}$ photons keV$^{-1}$ cm$^{-2}$ s$^{-1}$ at 1~keV } 
\end{tabular}
\end{table*}

Extensive work has been done on the degeneracy between the velocity broadening and the column density due to saturation of the ISM X-ray absorption lines \citep{jue06,yao06,bre07,gat13a,gat13b,luo14,fan15,nic16c,luo18}. The main difficulty when studying such lines is that the Doppler b-parameter ($b=\sqrt{2}v_{turb}$) cannot be fully constrained, which can lead to under(over)estimation of the column densities \citep{dra11,nic16c}. It is possible to measure the turbulence broadening if multiple transitions of the same ion are detected, however such measurement requires spectra with very good statistics and very good signal to noise ratio of the lines involved. For example, \citet{gat18} measured the Doppler b-parameter using the brightest sources in their data sample, Cygnus~X-2 for the Galactic sources and PKS~2155-304 for the Extragalactic sources. In order to evaluate the impact of the b-parameter in our analysis we performed a fit of the ISM component including $v_{turb}$ as a free parameter and using the best-fit results shown in Table~\ref{tab_best_fit_ism} as the initial fitting parameters. Table~\ref{tab_best_fit_ism_vturb} shows the best-fit parameters obtained. While the best-fit statistic is similar, the velocity uncertainties are large and the uncertainties in the column densities are much larger in comparison with the fit without $v_{turb}$ as free parameter. This may indicate saturation of the absorption lines.  Since the best-fit statistics obtained are similar than those listed in Table~\ref{tab_best_fit_ism}, we decided to fix the $v_{turb}$ parameter in the forthcoming analysis to those values derived by \citet{gat18} in order to save computational time.

\begin{table*}
\small
\caption{Best-fit parameters for the ISM component, including $v_{turb}$ as free parameter.}\label{tab_best_fit_ism_vturb}
\begin{tabular}{llccccccccccccccccccc}
\hline
Model&Parameter&1ES~1028+511& 1ES~1553+113  & H2356-309 & Mrk~421 & PG~1116+215 & PKS~0558-504   \\ 
\hline
 {\tt IONeq}$^{a}$ & $N({\rm H})$-cold &$4.79^{+0.66}_{-0.69}$  &$ 4.78  \pm 0.15  $  &$ 1.92_{-0.25}^{+0.23} $  &$ 1.65_{-0.17}^{+0.18}  $    &$ 1.08 \pm 0.48 $  &$  3.93_{-0.25}^{+0.30}  $     \\
&$N({\rm H})$-warm &$ <0.87 $  &$3.80_{-0.80}^{+1.11}  $  &$ <1.13 $  &$ 0.21_{-0.12}^{+0.07}   $    &$ 7.68_{-4.85}^{+4.33}  $  &$ 3.11_{-1.45}^{+1.31} $     \\
&$v_{turb}$-cold/warm &$5^{+58}_{-4} $  &$52^{+70}_{-39}     $  &$107_{-90}^{+102}   $&$ 60_{-55}^{+230}    $    &$125_{-115}^{199}    $  &$ 141_{-130}^{+50}   $     \\
&$N({\rm H})$-hot &$ <0.89  $  &$  0.36_{-0.24}^{+0.22}  $  &$ 0.72_{-0.46}^{+0.61}$  &$0.57_{-0.05}^{+0.07}   $    &$  9.82_{-4.41}^{+5.04}  $  &$  0.31_{-0.29}^{+0.36} $     \\
&$v_{turb}$-hot &$ 146_{-145}^{+853}   $  &$  89_{-59}^{+127}   $  &$329_{-293}^{+415}   $&$ 68_{-51}^{+182}    $    &$48_{-37}^{+432}    $  &$88_{-66}^{+194}    $     \\
statistics &cstat/d.o.f. &$ 2735/2660  $  &$  3767/2694  $  &$  2813/2691  $  &$  3088/2686  $    &$  2807/2669  $  &$  2947/2694 $     \\
\hline
\multicolumn{7}{l}{$^{a}$ $N({\rm H})$-cold in units of $10^{20}$ cm$^{-2}$. $N({\rm H})$-warm and $N({\rm H})$-hot in units of $10^{19}$ cm$^{-2}$.  $v_{turb}$ in units of km s$^{-1}$}\\ 

\end{tabular}
\end{table*}

\subsection{Searching for WHIM and/or CGM absorption features}\label{sec_best_fit_all}

\begin{figure}  
\includegraphics[scale=0.43]{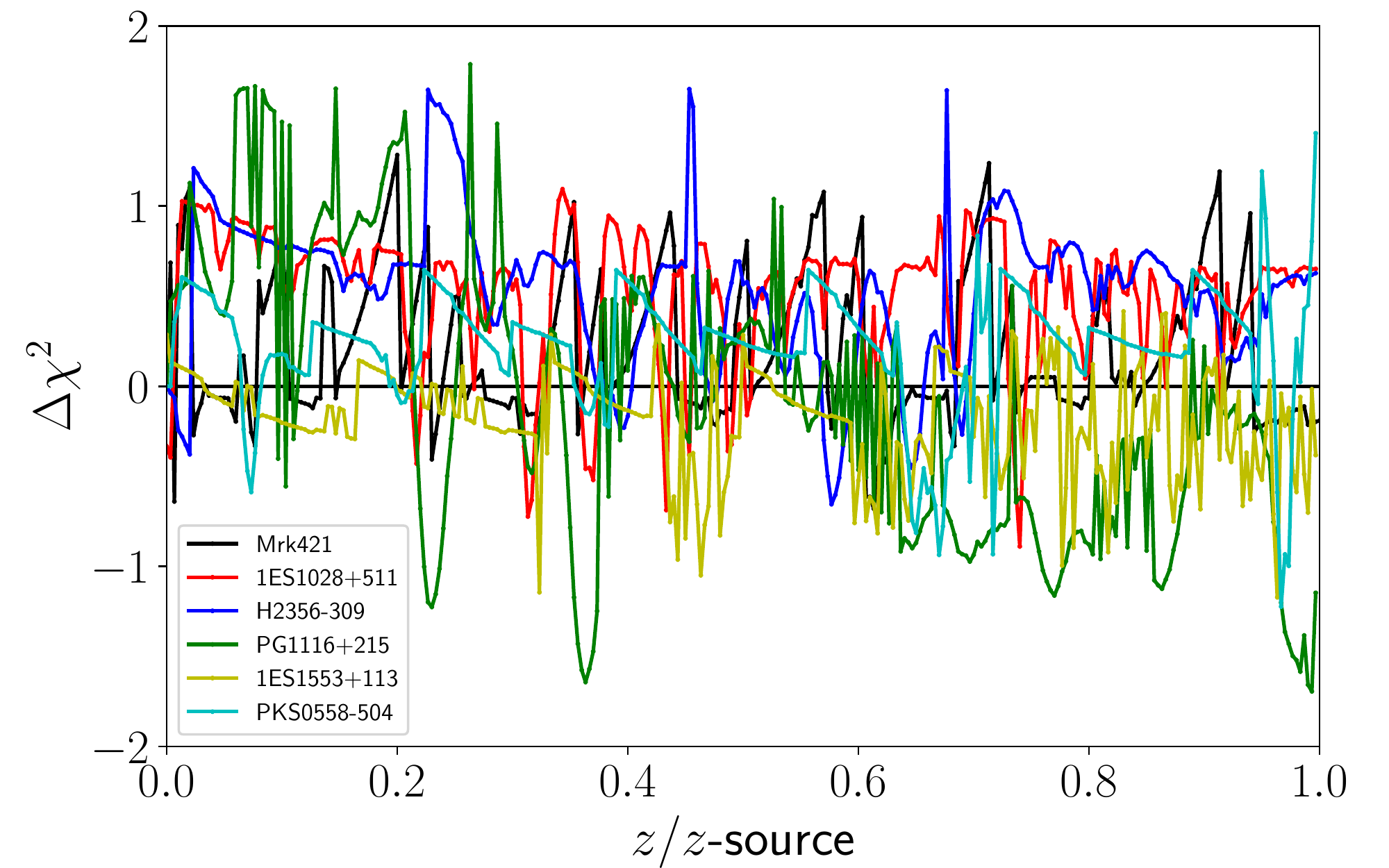}     
\caption{$\Delta \chi^{2} = \chi^{2}_{WHIM}-\chi^{2}_{ISM}$ variation when including the WHIM and/or CGM component compared with the best-fit shown in Table~\ref{tab_best_fit_ism}. The x-axis shows the distance to each source normalized (i.e. $z_{source}=1$).} \label{fig_delta_chisqr} 
\end{figure}

After fitting the ISM local contribution to the X-ray spectra, we model the WHIM and/or CGM X-ray absorption. It has been shown by simulations focused on individual galaxies \citep[e.g. ][]{dav10}, and by cosmological hydrodynamic simulations \citep[e.g. ][]{sti12} that the CGM is consistent with a plasma in CIE. Therefore, we have included a {\tt IONeq}$_{whim}$ component to model such environment, assuming solar abundances. It is important to note that our study is limited to the CIE component. For each source we search for a redshifted absorber by evaluating a grid of 300 redshift values between the observer ($z=0$) and the source distance. We impose the $T_{e}$  parameter in {\tt IONeq}$_{whim}$ to be in the range of values obtained in previous works (log(T) $5.4-6.4$, see Section~\ref{sec_dat}) while $N({\rm H})$ was let as an unconstrained free parameter. Figure~\ref{fig_delta_chisqr} shows the $\Delta \chi^{2} = \chi^{2}_{WHIM}-\chi^{2}_{ISM}$ value as function of the normalized redshift. We have found that, for all sources, the change in $\chi^{2}$ is not significant as we move along $z$. Cases where a slightly lower $\chi^{2}$ is obtained are due to small changes in the warm-hot column densities ($<1\%$), which is expected given the contour maps obtained for both components (Figure~\ref{fig_mcmc_par}). However, $N({\rm H})$-whim is always negligible.  That is, after modeling the ISM multicomponent we have not found a fourth component corresponding to the redshifted absorber in the line-of-sight to these sources.

\section{Discussion}\label{sec_disc} 
The identification of X-ray absorption features due to the WHIM has been a controversial topic in the last years. Most of claimed WHIM gas detection rely in the identification of a single X-ray absorption line located at a distance for which the presence of a filamentary structure is suspected or by doing a blind search. The accuracy of the atomic data included in the X-ray absorption models is crucial in the modeling of such absorption features. In this sense, the most up-to-date photoabsorption cross-sections are incorporated in the {\tt IONeq} model used in the analysis described in Section~\ref{sec_model} \citep{gat13a,gat13b,gor13,gat14,has14,gat15,gat18c,gat20b,leu20,gat21a}. Therefore, we are confident about the modeling of the local ISM X-ray absorption features, a crucial step before any attempt to search for WHIM absorption lines. Another important element in this type of analysis is to take into account the contribution from the multiple resonance lines, as well as ionic species, present at a given temperature since it can be not negligible. That is, any change in the ionic column density will affect simultaneously all absorption features related to the ion \citep[see for example][]{gat19a}.

  \begin{table} 
\caption{Relevant absorption line wavelengths.}\label{tab_lines}
\centering   
\begin{tabular}{lcccc}  
\hline
Source&Transition &   $\lambda$ (\AA) \\ 
 \hline
\hline
\\ 
ISM &N\,{\sc i} K$\alpha$ &$31.221 $    \\ 
WHIM &O\,{\sc vii} K$\alpha$ &$30.973 \pm 0.017^{a}$    \\
ISM &N\,{\sc ii} K$\alpha_{1}$ &$30.953 $    \\ 
ISM &N\,{\sc ii} K$\alpha_{2}$ &$30.931 $    \\ 
ISM &N\,{\sc ii} K$\alpha_{3}$ &$30.858 $    \\ 
ISM &N\,{\sc v} K$\alpha_{1}$ &$29.457 $    \\
WHIM &O\,{\sc vii} K$\alpha$ &$29.271 \pm 0.019^{a}$    \\ 
ISM &N\,{\sc v} K$\alpha_{2}$ &$29.128 $    \\ 
ISM &N\,{\sc vi} K$\alpha$ &$28.786 $    \\
WHIM &O\,{\sc vii} K$\alpha$ &$28.740\pm 0.038^{b}$    \\
ISM &N\,{\sc iii} K$\beta_{1}$ &$28.238 $    \\ 
ISM &O\,{\sc i} K$\beta$ &$22.887 $    \\ 
WHIM &O\,{\sc vii} K$\alpha$ &$ 22.290 \pm 0.017^{c}$    \\ 
ISM &O\,{\sc ii} K$\beta$ &$22.287 $    \\  
WHIM &O\,{\sc vii} K$\alpha$ &$22.184\pm 0.021^{d}$    \\
ISM &O\,{\sc vi} K$\alpha$ &$22.019 $    \\
ISM &O\,{\sc vi} K$\beta$ &$21.799 $    \\
ISM &O\,{\sc vii} K$\alpha$ &$21.601 $    \\
WHIM &O\,{\sc viii} K$\alpha$ &$21.187\pm 0.018^{e}$    \\
WHIM &O\,{\sc viii} K$\alpha$ &$20.696 \pm 0.017^{f}$    \\
\hline
\multicolumn{3}{l}{ $^{a}$ For 1ES 1553+113 \citep{nic18}.}   \\
\multicolumn{3}{l}{ $^{b}$ For 1ES~1028+511 \citep{nic05}.}\\
\multicolumn{3}{l}{ $^{c}$ For H2356-309 \citep{nev15}.} \\
\multicolumn{3}{l}{ $^{d}$ For Mrk~421 \citep{nic05b}.}   \\
\multicolumn{3}{l}{ $^{e}$ For PKS 0558-504 \citep{nic10}.}   \\
\multicolumn{3}{l}{ $^{f}$ For PG 1116+215 \citep{bon16}.}  
 \end{tabular}
\end{table}

Table~\ref{tab_lines} list the absorption lines associated to the WHIM that have been identified in previous works as well as the main ISM absorption lines located nearby (see zoom panels in Figure~\ref{fig_fits_flux}). The ISM line positions are taken from the {\sc xstar} database \citep{men21}. For Mrk~421, H2356-309 and PG~1116+215 the wavelength regions where WHIM absorption lines were supposedly located, are well modeled by the O\,{\sc ii}+O\,{\sc vi}+O\,{\sc vii} contribution from the local warm-hot ISM.  Our analysis of Mrk~421 support the conclusions from \citet{nic16b} about the absence of WHIM absorber in the line of sight through this source, with spectral absorption features being dominated entirely by the ISM. For 1ES~1553+113, the low residuals obtained by the N\,{\sc ii} contribution from the local warm ISM prevent us to identify O\,{\sc vii} redshifted absorption lines associated to the WHIM. For 1ES~1028+511, the wavelength region where a O\,{\sc vii} WHIM absorption line was identified in the past, here is well modeled with the N\,{\sc vii} contribution from the local hot ISM. In the case of PKS~0558-504, \citet{nic10} tentatively identified a WHIM absorption line with a statistical significance of $2.8\sigma$ and only upper limits for the line width were obtained ($<1200$ km/s). In our analysis, we have found that the global fit does not improve when including the WHIM component. 

\begin{figure}  
\includegraphics[width=0.48\textwidth]{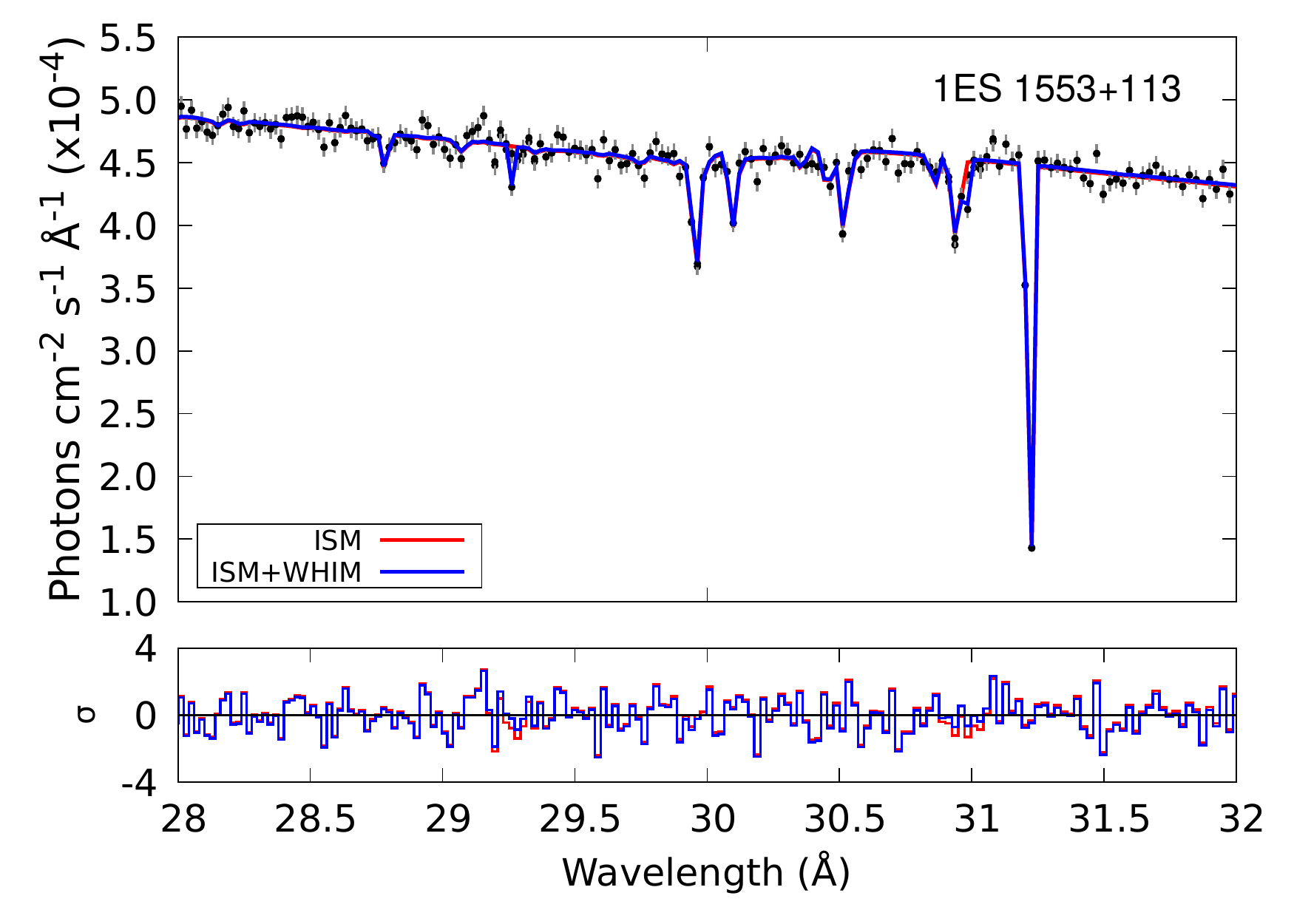}     
\caption{{\it Athena} simulation of 1ES~1553+113 in the N K-edge photoabsorption region withouth the WHIM components (red lines) and with the WHIM absorber (blue lines) as estimated by \citet{nic18}. The total exposure time is 100 ks and the flux is $\log F_{x}/$erg cm$^{-2}$ s$^{-1}$ $=-11.52$  in the 24-32 \AA\ wavelength range.} \label{fig_athena_sim} 
\end{figure}

It is clear that the complexity of the photoabsorption O and N K-edges makes it difficult to distinguish simultaneously absorbers from both environments WHIM and ISM, even more when considering that the ISM contribution is not always well constrained. Given the proximity in wavelength between lines, it is also possible that the claimed WHIM absorption features correspond to misidentified ISM lines. This points out the importance of modeling the multiphase ISM contribution before any attempt of searching for additional absorption features. Even for future X-ray high-resolution spectra mission, such as {\it Arcus} \citep{smi16}, {\it Athena} \citep{nan13} and the proposed Line Emission Mapper ({\it LEM}) probe \citep{lem22}, it is difficult to distinguish both absorbers, assuming that the WHIM absorbers are located at the claimed redshifts. For example, Figure~\ref{fig_athena_sim} shows a 100~ks {\it Athena} simulation of 1ES~1553+113 for the baseline configuration\footnote{\url{http://x-ifu-resources.irap.omp.eu/PUBLIC/RESPONSES/CC_CONFIGURATION/}}. For the multiphase ISM we used the parameters obtained in Section~\ref{sec_best_fit_all} while for the WHIM component we used the parameters reported by \citet{nic18}. Even for such moderate exposure, we have not found a significant improvement in the fit statistic when including the WHIM absorber at low-redshift. Given these difficulties, we recommend to search for WHIM absorption lines at redshifts where the multiphase ISM contribution is minimal (e.g. larger redshift). The development of cosmic filament catalogs in last years using spectroscopic redshift surveys (e.g. using SDSS data) will be very helpful as a reference for where to search for WHIM absorption features using X-ray high-resolution spectroscopy \citep[see for example][]{che16,mal20,ros20}. Naturally, the difficulty to perform such analysis will depend on the requirements for the background X-ray source: it should be bright and, ideally, the X-ray source spectra should be featureless (i.e. a blazar).

We note that our model assumes CIE for the WHIM absorber. While cosmological hydrodynamic simulations \citep[e.g.][]{dav10}, and higher resolution simulations focused on individual galaxies \citep[e.g.][]{sti12} point out the CIE conditions of the WHIM, it has been shown that interaction between the X-ray emissivity of this medium increases as consequence of interaction with the X-ray background photons \citep[e.g.][]{chu01}. In this sense, a future analysis will include photoionization effects in the WHIM absorption. Finally, while some of the WHIM absorption lines are expected to be saturated, such saturation will be less important than thermal line broadening for temperatures where the CIE ionization fractions peak \citep{wij20}.

There are some caveats in our analysis. First, we assume fixed (solar) abundance ratios and different abundance ratios will affect the column density measurements. Also, we assume that the broadening effect is independent on the line-of-sight direction. We are not including in our model intrinsic absorption associated to each extragalactic source. Our multi-temperature model consist of a single column density for each ISM phase, while varying optical depth values along the line-of-sight may require to define an optical depth distribution \citep[see for example][]{loc22}. Another limitation is that we fixed the temperatures of each ISM phase for all sources, thus ignoring local variations. Finally, uncertainties in the RGS effective areas at wavelengths corresponding to cool pixels in the dispersing detectors could accumulate in the process of combining the data.

\section{Conclusions}\label{sec_con}               
We analyzed six extragalactic sources searching for spectral signatures due to the WHIM and/or CGM using high-resolution X-ray spectra. We used the {\tt IONeq} model, which assumes collisional ionization equilibrium, in order to take into account the absorption features due to the cold, warm and hot components of the local ISM. We found acceptable fits for all sources with such simple model. We studied the parameter space of the best-fit results to analyze the confidence regions. After modeling the ISM component we included a {\tt IONeq} component to search for WHIM absorbers by creating a grid of redshifts between the observer and the source distance. We have not found a significant statistical improvement in any of the sources. This analysis shows that we can safely reject a successful detection of a redshifted CIE absorber towards the lines of sights to the sources. In all cases, the wavelength spectral region were WHIM absorption features were identified in previous analysis are well modeled by the local ISM. Our results point out the importance of modeling the X-ray absorption due to the multiphase ISM before any attempt to search for WHIM and/or CGM absorption features. Our simulation shows that the presence of the multiphase ISM absorption features prevents detection of low-redshift WHIM absorption features in the $>17$~\AA\ spectral region for moderate exposures using high-resolution spectra, including future observatories such as {\it Athena}. Finally, we recommend to search for redshifted absorption lines in spectral regions where the multiphase ISM contribution is minimal.

\subsection*{Data availability}
The observations analyzed in this article are available in the {\it XMM-Newton} Science Archive (XSA\footnote{\url{http://xmm.esac.esa.int/xsa/}}).

\bibliographystyle{mnras}

\begin{thebibliography}{}
\makeatletter
\relax
\def\mn@urlcharsother{\let\do\@makeother \do\$\do\&\do\#\do\^\do\_\do\%\do\~}
\def\mn@doi{\begingroup\mn@urlcharsother \@ifnextchar [ {\mn@doi@}
  {\mn@doi@[]}}
\def\mn@doi@[#1]#2{\def\@tempa{#1}\ifx\@tempa\@empty \href
  {http://dx.doi.org/#2} {doi:#2}\else \href {http://dx.doi.org/#2} {#1}\fi
  \endgroup}
\def\mn@eprint#1#2{\mn@eprint@#1:#2::\@nil}
\def\mn@eprint@arXiv#1{\href {http://arxiv.org/abs/#1} {{\tt arXiv:#1}}}
\def\mn@eprint@dblp#1{\href {http://dblp.uni-trier.de/rec/bibtex/#1.xml}
  {dblp:#1}}
\def\mn@eprint@#1:#2:#3:#4\@nil{\def\@tempa {#1}\def\@tempb {#2}\def\@tempc
  {#3}\ifx \@tempc \@empty \let \@tempc \@tempb \let \@tempb \@tempa \fi \ifx
  \@tempb \@empty \def\@tempb {arXiv}\fi \@ifundefined
  {mn@eprint@\@tempb}{\@tempb:\@tempc}{\expandafter \expandafter \csname
  mn@eprint@\@tempb\endcsname \expandafter{\@tempc}}}

\bibitem[\protect\citeauthoryear{{Abramowski} et~al.,}{{Abramowski}
  et~al.}{2015}]{hes15}
{Abramowski} A.,  et~al., 2015, \mn@doi [\apj] {10.1088/0004-637X/802/1/65},
  \href {https://ui.adsabs.harvard.edu/abs/2015ApJ...802...65A} {802, 65}

\bibitem[\protect\citeauthoryear{{Arnaud}}{{Arnaud}}{1996}]{arn96}
{Arnaud} K.~A.,  1996, in {Jacoby} G.~H.,  {Barnes} J.,  eds,  Astronomical
  Society of the Pacific Conference Series Vol. 101, Astronomical Data Analysis
  Software and Systems V. p.~17

\bibitem[\protect\citeauthoryear{{Bonamente}, {Nevalainen}, {Tilton},
  {Liivam{\"a}gi}, {Tempel}, {Hein{\"a}m{\"a}ki}  \& {Fang}}{{Bonamente}
  et~al.}{2016}]{bon16}
{Bonamente} M.,  {Nevalainen} J.,  {Tilton} E.,  {Liivam{\"a}gi} J.,  {Tempel}
  E.,  {Hein{\"a}m{\"a}ki} P.,   {Fang} T.,  2016, \mn@doi [\mnras]
  {10.1093/mnras/stw285}, \href
  {https://ui.adsabs.harvard.edu/abs/2016MNRAS.457.4236B} {457, 4236}

\bibitem[\protect\citeauthoryear{{Bregman} \& {Lloyd-Davies}}{{Bregman} \&
  {Lloyd-Davies}}{2007}]{bre07}
{Bregman} J.~N.,  {Lloyd-Davies} E.~J.,  2007, \mn@doi [\apj] {10.1086/521321},
  \href {https://ui.adsabs.harvard.edu/abs/2007ApJ...669..990B} {669, 990}

\bibitem[\protect\citeauthoryear{{Buote}, {Zappacosta}, {Fang}, {Humphrey},
  {Gastaldello}  \& {Tagliaferri}}{{Buote} et~al.}{2009}]{buo09}
{Buote} D.~A.,  {Zappacosta} L.,  {Fang} T.,  {Humphrey} P.~J.,  {Gastaldello}
  F.,   {Tagliaferri} G.,  2009, \mn@doi [\apj] {10.1088/0004-637X/695/2/1351},
  \href {https://ui.adsabs.harvard.edu/abs/2009ApJ...695.1351B} {695, 1351}

\bibitem[\protect\citeauthoryear{{Chen}, {Ho}, {Brinkmann}, {Freeman},
  {Genovese}, {Schneider}  \& {Wasserman}}{{Chen} et~al.}{2016}]{che16}
{Chen} Y.-C.,  {Ho} S.,  {Brinkmann} J.,  {Freeman} P.~E.,  {Genovese} C.~R.,
  {Schneider} D.~P.,   {Wasserman} L.,  2016, \mn@doi [\mnras]
  {10.1093/mnras/stw1554}, \href
  {https://ui.adsabs.harvard.edu/abs/2016MNRAS.461.3896C} {461, 3896}

\bibitem[\protect\citeauthoryear{{Churazov}, {Gilfanov}, {Forman}  \&
  {Jones}}{{Churazov} et~al.}{1996}]{chu96}
{Churazov} E.,  {Gilfanov} M.,  {Forman} W.,   {Jones} C.,  1996, \mn@doi
  [\apj] {10.1086/177997}, \href
  {http://adsabs.harvard.edu/abs/1996ApJ...471..673C} {471, 673}

\bibitem[\protect\citeauthoryear{{Churazov}, {Haehnelt}, {Kotov}  \&
  {Sunyaev}}{{Churazov} et~al.}{2001}]{chu01}
{Churazov} E.,  {Haehnelt} M.,  {Kotov} O.,   {Sunyaev} R.,  2001, \mn@doi
  [\mnras] {10.1046/j.1365-8711.2001.04090.x}, \href
  {https://ui.adsabs.harvard.edu/abs/2001MNRAS.323...93C} {323, 93}

\bibitem[\protect\citeauthoryear{{Crook}, {Huchra}, {Martimbeau}, {Masters},
  {Jarrett}  \& {Macri}}{{Crook} et~al.}{2007}]{cro07}
{Crook} A.~C.,  {Huchra} J.~P.,  {Martimbeau} N.,  {Masters} K.~L.,  {Jarrett}
  T.,   {Macri} L.~M.,  2007, \mn@doi [\apj] {10.1086/510201}, \href
  {https://ui.adsabs.harvard.edu/abs/2007ApJ...655..790C} {655, 790}

\bibitem[\protect\citeauthoryear{{Dav{\'e}}, {Spergel}, {Steinhardt}  \&
  {Wandelt}}{{Dav{\'e}} et~al.}{2001}]{dav01}
{Dav{\'e}} R.,  {Spergel} D.~N.,  {Steinhardt} P.~J.,   {Wandelt} B.~D.,  2001,
  \mn@doi [\apj] {10.1086/318417}, \href
  {https://ui.adsabs.harvard.edu/abs/2001ApJ...547..574D} {547, 574}

\bibitem[\protect\citeauthoryear{{Dav{\'e}}, {Oppenheimer}, {Katz}, {Kollmeier}
   \& {Weinberg}}{{Dav{\'e}} et~al.}{2010}]{dav10}
{Dav{\'e}} R.,  {Oppenheimer} B.~D.,  {Katz} N.,  {Kollmeier} J.~A.,
  {Weinberg} D.~H.,  2010, \mn@doi [\mnras] {10.1111/j.1365-2966.2010.17279.x},
  \href {https://ui.adsabs.harvard.edu/abs/2010MNRAS.408.2051D} {408, 2051}

\bibitem[\protect\citeauthoryear{{Donato}, {Ghisellini}, {Tagliaferri}  \&
  {Fossati}}{{Donato} et~al.}{2001}]{don01}
{Donato} D.,  {Ghisellini} G.,  {Tagliaferri} G.,   {Fossati} G.,  2001,
  \mn@doi [\aap] {10.1051/0004-6361:20010675}, \href
  {https://ui.adsabs.harvard.edu/abs/2001A&A...375..739D} {375, 739}

\bibitem[\protect\citeauthoryear{{Draine}}{{Draine}}{2011}]{dra11}
{Draine} B.~T.,  2011, {Physics of the Interstellar and Intergalactic Medium}

\bibitem[\protect\citeauthoryear{{Eracleous} \& {Halpern}}{{Eracleous} \&
  {Halpern}}{2004}]{era04}
{Eracleous} M.,  {Halpern} J.~P.,  2004, \mn@doi [\apjs] {10.1086/379823},
  \href {https://ui.adsabs.harvard.edu/abs/2004ApJS..150..181E} {150, 181}

\bibitem[\protect\citeauthoryear{{Fang}, {Buote}, {Humphrey}, {Canizares},
  {Zappacosta}, {Maiolino}, {Tagliaferri}  \& {Gastaldello}}{{Fang}
  et~al.}{2010}]{fan10}
{Fang} T.,  {Buote} D.~A.,  {Humphrey} P.~J.,  {Canizares} C.~R.,  {Zappacosta}
  L.,  {Maiolino} R.,  {Tagliaferri} G.,   {Gastaldello} F.,  2010, \mn@doi
  [\apj] {10.1088/0004-637X/714/2/1715}, \href
  {https://ui.adsabs.harvard.edu/abs/2010ApJ...714.1715F} {714, 1715}

\bibitem[\protect\citeauthoryear{{Fang}, {Buote}, {Bullock}  \& {Ma}}{{Fang}
  et~al.}{2015}]{fan15}
{Fang} T.,  {Buote} D.,  {Bullock} J.,   {Ma} R.,  2015, \mn@doi [\apjs]
  {10.1088/0067-0049/217/2/21}, \href
  {http://adsabs.harvard.edu/abs/2015ApJS..217...21F} {217, 21}

\bibitem[\protect\citeauthoryear{{Garc{\'{\i}}a}, {Mendoza}, {Bautista},
  {Gorczyca}, {Kallman}  \& {Palmeri}}{{Garc{\'{\i}}a} et~al.}{2005}]{gar05}
{Garc{\'{\i}}a} J.,  {Mendoza} C.,  {Bautista} M.~A.,  {Gorczyca} T.~W.,
  {Kallman} T.~R.,   {Palmeri} P.,  2005, \mn@doi [\apjs] {10.1086/428712},
  \href {http://adsabs.harvard.edu/abs/2005ApJS..158...68G} {158, 68}

\bibitem[\protect\citeauthoryear{{Garc{\'{\i}}a} et~al.,}{{Garc{\'{\i}}a}
  et~al.}{2009}]{gar09a}
{Garc{\'{\i}}a} J.,  et~al., 2009, \mn@doi [\apjs]
  {10.1088/0067-0049/185/2/477}, \href
  {http://adsabs.harvard.edu/abs/2009ApJS..185..477G} {185, 477}

\bibitem[\protect\citeauthoryear{{Gatuzz} \& {Churazov}}{{Gatuzz} \&
  {Churazov}}{2018}]{gat18}
{Gatuzz} E.,  {Churazov} E.,  2018, \mn@doi [\mnras] {10.1093/mnras/stx2776},
  \href {http://adsabs.harvard.edu/abs/2018MNRAS.474..696G} {474, 696}

\bibitem[\protect\citeauthoryear{{Gatuzz} et~al.,}{{Gatuzz}
  et~al.}{2013a}]{gat13a}
{Gatuzz} E.,  et~al., 2013a, \mn@doi [\apj] {10.1088/0004-637X/768/1/60}, \href
  {http://adsabs.harvard.edu/abs/2013ApJ...768...60G} {768, 60}

\bibitem[\protect\citeauthoryear{{Gatuzz} et~al.,}{{Gatuzz}
  et~al.}{2013b}]{gat13b}
{Gatuzz} E.,  et~al., 2013b, \mn@doi [\apj] {10.1088/0004-637X/778/1/83}, \href
  {http://adsabs.harvard.edu/abs/2013ApJ...778...83G} {778, 83}

\bibitem[\protect\citeauthoryear{{Gatuzz}, {Garc{\'{\i}}a}, {Mendoza},
  {Kallman}, {Bautista}  \& {Gorczyca}}{{Gatuzz} et~al.}{2014}]{gat14}
{Gatuzz} E.,  {Garc{\'{\i}}a} J.,  {Mendoza} C.,  {Kallman} T.~R.,  {Bautista}
  M.~A.,   {Gorczyca} T.~W.,  2014, \mn@doi [\apj]
  {10.1088/0004-637X/790/2/131}, \href
  {http://adsabs.harvard.edu/abs/2014ApJ...790..131G} {790, 131}

\bibitem[\protect\citeauthoryear{{Gatuzz}, {Garc{\'{\i}}a}, {Kallman},
  {Mendoza}  \& {Gorczyca}}{{Gatuzz} et~al.}{2015}]{gat15}
{Gatuzz} E.,  {Garc{\'{\i}}a} J.,  {Kallman} T.~R.,  {Mendoza} C.,   {Gorczyca}
  T.~W.,  2015, \mn@doi [\apj] {10.1088/0004-637X/800/1/29}, \href
  {http://adsabs.harvard.edu/abs/2015ApJ...800...29G} {800, 29}

\bibitem[\protect\citeauthoryear{{Gatuzz}, {Ness}, {Gorczyca}, {Hasoglu},
  {Kallman}  \& {Garc{\'\i}a}}{{Gatuzz} et~al.}{2018}]{gat18c}
{Gatuzz} E.,  {Ness} J.~U.,  {Gorczyca} T.~W.,  {Hasoglu} M.~F.,  {Kallman}
  T.~R.,   {Garc{\'\i}a} J.~A.,  2018, \mn@doi [\mnras]
  {10.1093/mnras/sty1517}, \href
  {https://ui.adsabs.harvard.edu/abs/2018MNRAS.479.2457G} {479, 2457}

\bibitem[\protect\citeauthoryear{{Gatuzz}, {Garc{\'\i}a}  \&
  {Kallman}}{{Gatuzz} et~al.}{2019}]{gat19a}
{Gatuzz} E.,  {Garc{\'\i}a} J.~A.,   {Kallman} T.~R.,  2019, \mn@doi [\mnras]
  {10.1093/mnrasl/sly223}, \href
  {https://ui.adsabs.harvard.edu/abs/2019MNRAS.483L..75G} {483, L75}

\bibitem[\protect\citeauthoryear{{Gatuzz}, {D{\'\i}az Trigo}, {Miller-Jones}
  \& {Migliari}}{{Gatuzz} et~al.}{2020a}]{gat20}
{Gatuzz} E.,  {D{\'\i}az Trigo} M.,  {Miller-Jones} J.~C.~A.,   {Migliari} S.,
  2020a, \mn@doi [\mnras] {10.1093/mnras/stz3385}, \href
  {https://ui.adsabs.harvard.edu/abs/2020MNRAS.491.4857G} {491, 4857}

\bibitem[\protect\citeauthoryear{{Gatuzz}, {Gorczyca}, {Hasoglu}, {Schulz},
  {Corrales}  \& {Mendoza}}{{Gatuzz} et~al.}{2020b}]{gat20b}
{Gatuzz} E.,  {Gorczyca} T.~W.,  {Hasoglu} M.~F.,  {Schulz} N.~S.,  {Corrales}
  L.,   {Mendoza} C.,  2020b, \mn@doi [\mnras] {10.1093/mnrasl/slaa119}, \href
  {https://ui.adsabs.harvard.edu/abs/2020MNRAS.498L..20G} {498, L20}

\bibitem[\protect\citeauthoryear{{Gatuzz}, {Garc{\'\i}a}  \&
  {Kallman}}{{Gatuzz} et~al.}{2021}]{gat21a}
{Gatuzz} E.,  {Garc{\'\i}a} J.~A.,   {Kallman} T.~R.,  2021, arXiv e-prints,
  \href {https://ui.adsabs.harvard.edu/abs/2021arXiv210411256G} {p.
  arXiv:2104.11256}

\bibitem[\protect\citeauthoryear{{Ghosh}, {Dewangan}  \&
  {Raychaudhuri}}{{Ghosh} et~al.}{2016}]{gho16}
{Ghosh} R.,  {Dewangan} G.~C.,   {Raychaudhuri} B.,  2016, \mn@doi [\mnras]
  {10.1093/mnras/stv2682}, \href
  {https://ui.adsabs.harvard.edu/abs/2016MNRAS.456..554G} {456, 554}

\bibitem[\protect\citeauthoryear{{Gliozzi}, {Papadakis}, {Grupe}, {Brinkmann},
  {Raeth}  \& {Kedziora-Chudczer}}{{Gliozzi} et~al.}{2010}]{gli10}
{Gliozzi} M.,  {Papadakis} I.~E.,  {Grupe} D.,  {Brinkmann} W.~P.,  {Raeth} C.,
    {Kedziora-Chudczer} L.,  2010, \mn@doi [\apj]
  {10.1088/0004-637X/717/2/1243}, \href
  {https://ui.adsabs.harvard.edu/abs/2010ApJ...717.1243G} {717, 1243}

\bibitem[\protect\citeauthoryear{{Gliozzi}, {Papadakis}, {Grupe}, {Brinkmann}
  \& {R{\"a}th}}{{Gliozzi} et~al.}{2013}]{gli13}
{Gliozzi} M.,  {Papadakis} I.~E.,  {Grupe} D.,  {Brinkmann} W.~P.,   {R{\"a}th}
  C.,  2013, \mn@doi [\mnras] {10.1093/mnras/stt848}, \href
  {https://ui.adsabs.harvard.edu/abs/2013MNRAS.433.1709G} {433, 1709}

\bibitem[\protect\citeauthoryear{{Gorczyca} et~al.,}{{Gorczyca}
  et~al.}{2013}]{gor13}
{Gorczyca} T.~W.,  et~al., 2013, \mn@doi [\apj] {10.1088/0004-637X/779/1/78},
  \href {http://adsabs.harvard.edu/abs/2013ApJ...779...78G} {779, 78}

\bibitem[\protect\citeauthoryear{{Grevesse} \& {Sauval}}{{Grevesse} \&
  {Sauval}}{1998}]{gre98}
{Grevesse} N.,  {Sauval} A.~J.,  1998, \mn@doi [\ssr]
  {10.1023/A:1005161325181}, \href
  {http://adsabs.harvard.edu/abs/1998SSRv...85..161G} {85, 161}

\bibitem[\protect\citeauthoryear{{Han}, {Tian}, {Ji}, {Liu}  \& {Zhou}}{{Han}
  et~al.}{2014}]{han14}
{Han} X.~L.,  {Tian} Q.,  {Ji} T.,  {Liu} W.,   {Zhou} H.,  2014, in American
  Astronomical Society Meeting Abstracts \#224. p. 417.05

\bibitem[\protect\citeauthoryear{{Haso{\u g}lu}, {Abdel-Naby}, {Gatuzz},
  {Garc\'ia}, {Kallman}, {Mendoza}  \& {Gorczyca}}{{Haso{\u g}lu}
  et~al.}{2014}]{has14}
{Haso{\u g}lu} M.~F.,  {Abdel-Naby} S.~A.,  {Gatuzz} E.,  {Garc\'ia} J.,
  {Kallman} T.~R.,  {Mendoza} C.,   {Gorczyca} T.~W.,  2014, The Astrophysical
  Journal Supplement Series, 214, 8

\bibitem[\protect\citeauthoryear{{Joachimi}, {Gatuzz}, {Garc{\'\i}a}  \&
  {Kallman}}{{Joachimi} et~al.}{2016}]{joa16}
{Joachimi} K.,  {Gatuzz} E.,  {Garc{\'\i}a} J.~A.,   {Kallman} T.~R.,  2016,
  \mn@doi [\mnras] {10.1093/mnras/stw1371}, \href
  {https://ui.adsabs.harvard.edu/abs/2016MNRAS.461..352J} {461, 352}

\bibitem[\protect\citeauthoryear{{Johnson} et~al.,}{{Johnson}
  et~al.}{2019}]{joh19}
{Johnson} S.~D.,  et~al., 2019, \mn@doi [\apjl] {10.3847/2041-8213/ab479a},
  \href {https://ui.adsabs.harvard.edu/abs/2019ApJ...884L..31J} {884, L31}

\bibitem[\protect\citeauthoryear{{Juett}, {Schulz}, {Chakrabarty}  \&
  {Gorczyca}}{{Juett} et~al.}{2006}]{jue06}
{Juett} A.~M.,  {Schulz} N.~S.,  {Chakrabarty} D.,   {Gorczyca} T.~W.,  2006,
  \mn@doi [\apj] {10.1086/506189}, \href
  {http://adsabs.harvard.edu/abs/2006ApJ...648.1066J} {648, 1066}

\bibitem[\protect\citeauthoryear{{Kaastra}, {Werner}, {Herder}, {Paerels}, {de
  Plaa}, {Rasmussen}  \& {de Vries}}{{Kaastra} et~al.}{2006}]{kaa06}
{Kaastra} J.~S.,  {Werner} N.,  {Herder} J.~W.~A.~d.,  {Paerels} F.~B.~S.,  {de
  Plaa} J.,  {Rasmussen} A.~P.,   {de Vries} C.~P.,  2006, \mn@doi [\apj]
  {10.1086/507835}, \href
  {https://ui.adsabs.harvard.edu/abs/2006ApJ...652..189K} {652, 189}

\bibitem[\protect\citeauthoryear{{Kraft} et~al.,}{{Kraft} et~al.}{2022}]{lem22}
{Kraft} R.,  et~al., 2022, arXiv e-prints, \href
  {https://ui.adsabs.harvard.edu/abs/2022arXiv221109827K} {p. arXiv:2211.09827}

\bibitem[\protect\citeauthoryear{{Kurtanidze}, {Nikolashvili}, {Kapanadze},
  {Kimeridze}, {Sigua}, {Urushadze}  \& {Goderidze}}{{Kurtanidze}
  et~al.}{2004}]{kur04}
{Kurtanidze} O.~M.,  {Nikolashvili} M.~G.,  {Kapanadze} B.~Z.,  {Kimeridze}
  G.~N.,  {Sigua} L.~A.,  {Urushadze} T.~V.,   {Goderidze} E.~K.,  2004,
  \mn@doi [Nuclear Physics B Proceedings Supplements]
  {10.1016/j.nuclphysbps.2004.04.031}, \href
  {https://ui.adsabs.harvard.edu/abs/2004NuPhS.132..193K} {132, 193}

\bibitem[\protect\citeauthoryear{{Lehner}, {Savage}, {Richter}, {Sembach},
  {Tripp}  \& {Wakker}}{{Lehner} et~al.}{2007}]{leh07}
{Lehner} N.,  {Savage} B.~D.,  {Richter} P.,  {Sembach} K.~R.,  {Tripp} T.~M.,
   {Wakker} B.~P.,  2007, \mn@doi [\apj] {10.1086/511749}, \href
  {https://ui.adsabs.harvard.edu/abs/2007ApJ...658..680L} {658, 680}

\bibitem[\protect\citeauthoryear{{Leutenegger} et~al.,}{{Leutenegger}
  et~al.}{2020}]{leu20}
{Leutenegger} M.~A.,  et~al., 2020, \mn@doi [\prl]
  {10.1103/PhysRevLett.125.243001}, \href
  {https://ui.adsabs.harvard.edu/abs/2020PhRvL.125x3001L} {125, 243001}

\bibitem[\protect\citeauthoryear{{Locatelli}, {Ponti}  \&
  {Bianchi}}{{Locatelli} et~al.}{2022}]{loc22}
{Locatelli} N.,  {Ponti} G.,   {Bianchi} S.,  2022, \mn@doi [\aap]
  {10.1051/0004-6361/202142655}, \href
  {https://ui.adsabs.harvard.edu/abs/2022A&A...659A.118L} {659, A118}

\bibitem[\protect\citeauthoryear{{Luo} \& {Fang}}{{Luo} \&
  {Fang}}{2014}]{luo14}
{Luo} Y.,  {Fang} T.,  2014, \mn@doi [\apj] {10.1088/0004-637X/780/2/170},
  \href {http://adsabs.harvard.edu/abs/2014ApJ...780..170L} {780, 170}

\bibitem[\protect\citeauthoryear{{Luo}, {Fang}  \& {Ma}}{{Luo}
  et~al.}{2018}]{luo18}
{Luo} Y.,  {Fang} T.,   {Ma} R.,  2018, \mn@doi [\apjs]
  {10.3847/1538-4365/aab270}, \href
  {https://ui.adsabs.harvard.edu/abs/2018ApJS..235...28L} {235, 28}

\bibitem[\protect\citeauthoryear{{Malavasi}, {Aghanim}, {Douspis}, {Tanimura}
  \& {Bonjean}}{{Malavasi} et~al.}{2020}]{mal20}
{Malavasi} N.,  {Aghanim} N.,  {Douspis} M.,  {Tanimura} H.,   {Bonjean} V.,
  2020, \mn@doi [\aap] {10.1051/0004-6361/202037647}, \href
  {https://ui.adsabs.harvard.edu/abs/2020A&A...642A..19M} {642, A19}

\bibitem[\protect\citeauthoryear{{Medvedev}, {Khabibullin}, {Sazonov},
  {Churazov}  \& {Tsygankov}}{{Medvedev} et~al.}{2018}]{med18}
{Medvedev} P.~S.,  {Khabibullin} I.~I.,  {Sazonov} S.~Y.,  {Churazov} E.~M.,
  {Tsygankov} S.~S.,  2018, \mn@doi [Astronomy Letters]
  {10.1134/S1063773718060038}, \href
  {https://ui.adsabs.harvard.edu/abs/2018AstL...44..390M} {44, 390}

\bibitem[\protect\citeauthoryear{{Mendoza} et~al.,}{{Mendoza}
  et~al.}{2021}]{men21}
{Mendoza} C.,  et~al., 2021, \mn@doi [Atoms] {10.3390/atoms9010012}, \href
  {https://ui.adsabs.harvard.edu/abs/2021Atoms...9...12M} {9, 12}

\bibitem[\protect\citeauthoryear{{Miller} et~al.,}{{Miller}
  et~al.}{2016}]{mil16d}
{Miller} J.~M.,  et~al., 2016, \mn@doi [\apjl] {10.3847/2041-8205/821/1/L9},
  \href {https://ui.adsabs.harvard.edu/abs/2016ApJ...821L...9M} {821, L9}

\bibitem[\protect\citeauthoryear{{Nandra} et~al.,}{{Nandra}
  et~al.}{2013}]{nan13}
{Nandra} K.,  et~al., 2013, arXiv e-prints, \href
  {https://ui.adsabs.harvard.edu/abs/2013arXiv1306.2307N} {p. arXiv:1306.2307}

\bibitem[\protect\citeauthoryear{{Nelson} et~al.,}{{Nelson}
  et~al.}{2018}]{nel18}
{Nelson} D.,  et~al., 2018, \mn@doi [\mnras] {10.1093/mnras/sty656}, \href
  {https://ui.adsabs.harvard.edu/abs/2018MNRAS.477..450N} {477, 450}

\bibitem[\protect\citeauthoryear{{Nevalainen} et~al.,}{{Nevalainen}
  et~al.}{2015}]{nev15}
{Nevalainen} J.,  et~al., 2015, \mn@doi [\aap] {10.1051/0004-6361/201526443},
  \href {https://ui.adsabs.harvard.edu/abs/2015A&A...583A.142N} {583, A142}

\bibitem[\protect\citeauthoryear{{Nicastro}, {Elvis}, {Fiore}  \&
  {Mathur}}{{Nicastro} et~al.}{2005a}]{nic05}
{Nicastro} F.,  {Elvis} M.,  {Fiore} F.,   {Mathur} S.,  2005a, \mn@doi
  [Advances in Space Research] {10.1016/j.asr.2005.01.053}, \href
  {https://ui.adsabs.harvard.edu/abs/2005AdSpR..36..721N} {36, 721}

\bibitem[\protect\citeauthoryear{{Nicastro} et~al.,}{{Nicastro}
  et~al.}{2005b}]{nic05b}
{Nicastro} F.,  et~al., 2005b, \mn@doi [\nat] {10.1038/nature03245}, \href
  {https://ui.adsabs.harvard.edu/abs/2005Natur.433..495N} {433, 495}

\bibitem[\protect\citeauthoryear{{Nicastro} et~al.,}{{Nicastro}
  et~al.}{2005c}]{nic05a}
{Nicastro} F.,  et~al., 2005c, \mn@doi [\apj] {10.1086/431270}, \href
  {https://ui.adsabs.harvard.edu/abs/2005ApJ...629..700N} {629, 700}

\bibitem[\protect\citeauthoryear{{Nicastro}, {Krongold}, {Fields},
  {Conciatore}, {Zappacosta}, {Elvis}, {Mathur}  \& {Papadakis}}{{Nicastro}
  et~al.}{2010}]{nic10}
{Nicastro} F.,  {Krongold} Y.,  {Fields} D.,  {Conciatore} M.~L.,  {Zappacosta}
  L.,  {Elvis} M.,  {Mathur} S.,   {Papadakis} I.,  2010, \mn@doi [\apj]
  {10.1088/0004-637X/715/2/854}, \href
  {https://ui.adsabs.harvard.edu/abs/2010ApJ...715..854N} {715, 854}

\bibitem[\protect\citeauthoryear{{Nicastro}, {Senatore}, {Gupta}, {Mathur},
  {Krongold}, {Elvis}  \& {Piro}}{{Nicastro} et~al.}{2016a}]{nic16b}
{Nicastro} F.,  {Senatore} F.,  {Gupta} A.,  {Mathur} S.,  {Krongold} Y.,
  {Elvis} M.,   {Piro} L.,  2016a, \mn@doi [\mnras] {10.1093/mnrasl/slw022},
  \href {http://adsabs.harvard.edu/abs/2016MNRAS.458L.123N} {458, L123}

\bibitem[\protect\citeauthoryear{{Nicastro}, {Senatore}, {Krongold}, {Mathur}
  \& {Elvis}}{{Nicastro} et~al.}{2016b}]{nic16c}
{Nicastro} F.,  {Senatore} F.,  {Krongold} Y.,  {Mathur} S.,   {Elvis} M.,
  2016b, \mn@doi [\apjl] {10.3847/2041-8205/828/1/L12}, \href
  {http://adsabs.harvard.edu/abs/2016ApJ...828L..12N} {828, L12}

\bibitem[\protect\citeauthoryear{{Nicastro} et~al.,}{{Nicastro}
  et~al.}{2018}]{nic18}
{Nicastro} F.,  et~al., 2018, \mn@doi [\nat] {10.1038/s41586-018-0204-1}, \href
  {https://ui.adsabs.harvard.edu/abs/2018Natur.558..406N} {558, 406}

\bibitem[\protect\citeauthoryear{{Palmeri}, {Quinet}, {Mendoza}, {Bautista},
  {Witthoeft}  \& {Kallman}}{{Palmeri} et~al.}{2016}]{pal16}
{Palmeri} P.,  {Quinet} P.,  {Mendoza} C.,  {Bautista} M.~A.,  {Witthoeft}
  M.~C.,   {Kallman} T.~R.,  2016, \mn@doi [Astron. Astrophys.]
  {10.1051/0004-6361/201628457}, \href
  {https://ui.adsabs.harvard.edu/abs/2016A&A...589A.137P} {589, A137}

\bibitem[\protect\citeauthoryear{{Porter} \& {Raychaudhury}}{{Porter} \&
  {Raychaudhury}}{2005}]{por05}
{Porter} S.~C.,  {Raychaudhury} S.,  2005, \mn@doi [\mnras]
  {10.1111/j.1365-2966.2005.09688.x}, \href
  {https://ui.adsabs.harvard.edu/abs/2005MNRAS.364.1387P} {364, 1387}

\bibitem[\protect\citeauthoryear{{Rasmussen}, {Kahn}, {Paerels}, {Herder},
  {Kaastra}  \& {de Vries}}{{Rasmussen} et~al.}{2007}]{ras07}
{Rasmussen} A.~P.,  {Kahn} S.~M.,  {Paerels} F.,  {Herder} J. W.~d.,  {Kaastra}
  J.,   {de Vries} C.,  2007, \mn@doi [\apj] {10.1086/509865}, \href
  {https://ui.adsabs.harvard.edu/abs/2007ApJ...656..129R} {656, 129}

\bibitem[\protect\citeauthoryear{{Rost}, {Stasyszyn}, {Pereyra}  \&
  {Mart{\'\i}nez}}{{Rost} et~al.}{2020}]{ros20}
{Rost} A.,  {Stasyszyn} F.,  {Pereyra} L.,   {Mart{\'\i}nez} H.~J.,  2020,
  \mn@doi [\mnras] {10.1093/mnras/staa320}, \href
  {https://ui.adsabs.harvard.edu/abs/2020MNRAS.493.1936R} {493, 1936}

\bibitem[\protect\citeauthoryear{{Sembach}, {Tripp}, {Savage}  \&
  {Richter}}{{Sembach} et~al.}{2004}]{sem04}
{Sembach} K.~R.,  {Tripp} T.~M.,  {Savage} B.~D.,   {Richter} P.,  2004,
  \mn@doi [\apjs] {10.1086/425037}, \href
  {https://ui.adsabs.harvard.edu/abs/2004ApJS..155..351S} {155, 351}

\bibitem[\protect\citeauthoryear{{Shull}, {Smith}  \& {Danforth}}{{Shull}
  et~al.}{2012}]{shu12}
{Shull} J.~M.,  {Smith} B.~D.,   {Danforth} C.~W.,  2012, \mn@doi [\apj]
  {10.1088/0004-637X/759/1/23}, \href
  {https://ui.adsabs.harvard.edu/abs/2012ApJ...759...23S} {759, 23}

\bibitem[\protect\citeauthoryear{{Smith} et~al.,}{{Smith} et~al.}{2016}]{smi16}
{Smith} R.~K.,  et~al., 2016, in \procspie. p. 99054M,
  \mn@doi{10.1117/12.2231778}

\bibitem[\protect\citeauthoryear{{Steenbrugge}, {Nicastro}  \&
  {Elvis}}{{Steenbrugge} et~al.}{2006}]{ste06}
{Steenbrugge} K.~C.,  {Nicastro} F.,   {Elvis} M.,  2006, in {Wilson} A.,  ed.,
   ESA Special Publication Vol. 604, The X-ray Universe 2005. p.~751
  (\mn@eprint {arXiv} {astro-ph/0511234})

\bibitem[\protect\citeauthoryear{{Stinson} et~al.,}{{Stinson}
  et~al.}{2012}]{sti12}
{Stinson} G.~S.,  et~al., 2012, \mn@doi [\mnras]
  {10.1111/j.1365-2966.2012.21522.x}, \href
  {https://ui.adsabs.harvard.edu/abs/2012MNRAS.425.1270S} {425, 1270}

\bibitem[\protect\citeauthoryear{{Stolte} et~al.,}{{Stolte}
  et~al.}{1997}]{sto97}
{Stolte} W.~C.,  et~al., 1997, \mn@doi [Journal of Physics B Atomic Molecular
  Physics] {10.1088/0953-4075/30/20/012}, \href
  {http://adsabs.harvard.edu/abs/1997JPhB...30.4489S} {30, 4489}

\bibitem[\protect\citeauthoryear{{Tempel} et~al.,}{{Tempel}
  et~al.}{2014}]{tem14}
{Tempel} E.,  et~al., 2014, \mn@doi [\aap] {10.1051/0004-6361/201423585}, \href
  {https://ui.adsabs.harvard.edu/abs/2014A&A...566A...1T} {566, A1}

\bibitem[\protect\citeauthoryear{{Tilton}, {Danforth}, {Shull}  \&
  {Ross}}{{Tilton} et~al.}{2012}]{til12}
{Tilton} E.~M.,  {Danforth} C.~W.,  {Shull} J.~M.,   {Ross} T.~L.,  2012,
  \mn@doi [\apj] {10.1088/0004-637X/759/2/112}, \href
  {https://ui.adsabs.harvard.edu/abs/2012ApJ...759..112T} {759, 112}

\bibitem[\protect\citeauthoryear{{Tofflemire}, {Orio}, {Page}, {Osborne},
  {Ciroi}, {Cracco}, {Di Mille}  \& {Maxwell}}{{Tofflemire}
  et~al.}{2013}]{tof13}
{Tofflemire} B.~M.,  {Orio} M.,  {Page} K.~L.,  {Osborne} J.~P.,  {Ciroi} S.,
  {Cracco} V.,  {Di Mille} F.,   {Maxwell} M.,  2013, \mn@doi [\apj]
  {10.1088/0004-637X/779/1/22}, \href
  {https://ui.adsabs.harvard.edu/abs/2013ApJ...779...22T} {779, 22}

\bibitem[\protect\citeauthoryear{{Wijers}, {Schaye}  \& {Oppenheimer}}{{Wijers}
  et~al.}{2020}]{wij20}
{Wijers} N.~A.,  {Schaye} J.,   {Oppenheimer} B.~D.,  2020, \mn@doi [\mnras]
  {10.1093/mnras/staa2456}, \href
  {https://ui.adsabs.harvard.edu/abs/2020MNRAS.498..574W} {498, 574}

\bibitem[\protect\citeauthoryear{{Williams}, {Mulchaey}  \&
  {Kollmeier}}{{Williams} et~al.}{2013a}]{wil13a}
{Williams} R.~J.,  {Mulchaey} J.~S.,   {Kollmeier} J.~A.,  2013a, \mn@doi
  [\apjl] {10.1088/2041-8205/762/1/L10}, \href
  {https://ui.adsabs.harvard.edu/abs/2013ApJ...762L..10W} {762, L10}

\bibitem[\protect\citeauthoryear{{Williams}, {Mulchaey}  \&
  {Kollmeier}}{{Williams} et~al.}{2013b}]{will13}
{Williams} R.~J.,  {Mulchaey} J.~S.,   {Kollmeier} J.~A.,  2013b, \mn@doi
  [\apjl] {10.1088/2041-8205/762/1/L10}, \href
  {https://ui.adsabs.harvard.edu/abs/2013ApJ...762L..10W} {762, L10}

\bibitem[\protect\citeauthoryear{{Willingale}, {Starling}, {Beardmore},
  {Tanvir}  \& {O'Brien}}{{Willingale} et~al.}{2013}]{wil13}
{Willingale} R.,  {Starling} R.~L.~C.,  {Beardmore} A.~P.,  {Tanvir} N.~R.,
  {O'Brien} P.~T.,  2013, \mn@doi [\mnras] {10.1093/mnras/stt175}, \href
  {http://adsabs.harvard.edu/abs/2013MNRAS.431..394W} {431, 394}

\bibitem[\protect\citeauthoryear{{Witthoeft}, {Bautista}, {Mendoza}, {Kallman},
  {Palmeri}  \& {Quinet}}{{Witthoeft} et~al.}{2009}]{wit09}
{Witthoeft} M.~C.,  {Bautista} M.~A.,  {Mendoza} C.,  {Kallman} T.~R.,
  {Palmeri} P.,   {Quinet} P.,  2009, \mn@doi [\apjs]
  {10.1088/0067-0049/182/1/127}, \href
  {http://adsabs.harvard.edu/abs/2009ApJS..182..127W} {182, 127}

\bibitem[\protect\citeauthoryear{{Witthoeft}, {Garc{\'\i}a}, {Kallman},
  {Bautista}, {Mendoza}, {Palmeri}  \& {Quinet}}{{Witthoeft}
  et~al.}{2011a}]{wit11b}
{Witthoeft} M.~C.,  {Garc{\'\i}a} J.,  {Kallman} T.~R.,  {Bautista} M.~A.,
  {Mendoza} C.,  {Palmeri} P.,   {Quinet} P.,  2011a, \mn@doi [Astrophys. J.
  Suppl. Ser.] {10.1088/0067-0049/192/1/7}, \href
  {https://ui.adsabs.harvard.edu/abs/2011ApJS..192....7W} {192, 7}

\bibitem[\protect\citeauthoryear{{Witthoeft}, {Bautista}, {Garc{\'\i}a},
  {Kallman}, {Mendoza}, {Palmeri}  \& {Quinet}}{{Witthoeft}
  et~al.}{2011b}]{wit11a}
{Witthoeft} M.~C.,  {Bautista} M.~A.,  {Garc{\'\i}a} J.,  {Kallman} T.~R.,
  {Mendoza} C.,  {Palmeri} P.,   {Quinet} P.,  2011b, \mn@doi [Astrophys. J.
  Suppl. Ser.] {10.1088/0067-0049/196/1/7}, \href
  {https://ui.adsabs.harvard.edu/abs/2011ApJS..196....7W} {196, 7}

\bibitem[\protect\citeauthoryear{{Yao} \& {Wang}}{{Yao} \&
  {Wang}}{2006}]{yao06}
{Yao} Y.,  {Wang} Q.~D.,  2006, \mn@doi [\apj] {10.1086/500506}, \href
  {http://adsabs.harvard.edu/abs/2006ApJ...641..930Y} {641, 930}

\bibitem[\protect\citeauthoryear{{Yao}, {Shull}, {Wang}  \& {Cash}}{{Yao}
  et~al.}{2012}]{yao12}
{Yao} Y.,  {Shull} J.~M.,  {Wang} Q.~D.,   {Cash} W.,  2012, \mn@doi [\apj]
  {10.1088/0004-637X/746/2/166}, \href
  {https://ui.adsabs.harvard.edu/abs/2012ApJ...746..166Y} {746, 166}

\bibitem[\protect\citeauthoryear{{Zappacosta}, {Nicastro}, {Maiolino},
  {Tagliaferri}, {Buote}, {Fang}, {Humphrey}  \& {Gastaldello}}{{Zappacosta}
  et~al.}{2010}]{zap10}
{Zappacosta} L.,  {Nicastro} F.,  {Maiolino} R.,  {Tagliaferri} G.,  {Buote}
  D.~A.,  {Fang} T.,  {Humphrey} P.~J.,   {Gastaldello} F.,  2010, \mn@doi
  [\apj] {10.1088/0004-637X/717/1/74}, \href
  {https://ui.adsabs.harvard.edu/abs/2010ApJ...717...74Z} {717, 74}

\bibitem[\protect\citeauthoryear{{da Costa} et~al.,}{{da Costa}
  et~al.}{1994}]{dac94}
{da Costa} L.~N.,  et~al., 1994, \mn@doi [\apjl] {10.1086/187260}, \href
  {https://ui.adsabs.harvard.edu/abs/1994ApJ...424L...1D} {424, L1}

\makeatother
\end{thebibliography}

\appendix\label{sec_apx}

\section{Observation IDs}\label{sec_obsids}
Table~\ref{tab_obsids} list the {\it XMM-Newton} observation IDs analyzed in this work.

\begin{table*}
\caption{\label{tab_obsids}{\it XMM-Newton}  observation IDs analyzed in the present work.}
\centering
\small
\begin{tabular}{cccccccc}
\hline
\multicolumn{2}{c}{Mrk~421}& 1ES~1028+511 & 1ES~1553+113 & H2356-309    & PKS~0558-504 & PG~1116+215   \\
\hline
0099280101&0411081501&0094382701   &0790381401&0304080501 &0117500201&0111290401\\
0099280201&0411081601&0303720201   &0094380801&0304080601 &0117710601&0201940101\\
0136540101&0411081701&0303720301   &0656990101&0693500101 &0117710701&0201940201\\
0136540201&0411081901&0303720601   &0727780101&0722860101 &0120300801&0554380101\\
0136540301&0411082001&&0727780301&0722860201& 0125110101&0554380201\\
0136540401&0411082101&&0727780401&0722860301& 0129360201&0554380301\\
0136540601&0411082201&&0727780501&0722860401& 0137550201&&\\
0150498701&0411082301&&0761100101&0722860701& 0137550601&&\\
0153950601&0411082401&&0761100201& & 0555170201&&\\
0153950801&0411082501&&0761100301& & 0555170301&&\\
0158970101&0411082701&&0761100401& & 0555170401&&\\
0158970201&0411083201&&0761100701& & 0555170501&&\\
0158970701&0502030101&&0761101001& & 0555170601&&\\
0158971201&0510610101&&0790380501& &&&\\
0158971301&0510610201&&0790380601& &&&\\
0162960101&0560980101&&0790380801& &&&\\
0411080701&0560983301&&0790381001& &&&\\
0411080801&0656380101&&0790381501& &&&\\
0411081301&&&0810830101& &&&\\
0411081401&&&0810830201& &&&\\ 
\hline  
\end{tabular}
\end{table*}

\end{document}